\documentclass[fleqn]{aa}  

\usepackage{graphicx}
\usepackage{xcolor}
\usepackage{hyperref}
\usepackage{txfonts}
\usepackage{calligra}
\usepackage{calrsfs}
\usepackage[T1]{fontenc}
\usepackage[mathcal]{eucal}
\usepackage{longtable}
\newcommand{\Msun}{\, {\rm M}_{\odot}}
\newcommand{\Lsun}{\, {\rm L}_{\odot}}
\newcommand{\Rsun}{\, {\rm R}_{\odot}}
\newcommand{\Msunpc}{{\rm M}_{\odot}\, {\rm pc}}

\begin{document}


    \title{Imaging dark matter at the smallest scales with $z\approx1$ lensed stars}
  
  \titlerunning{3M-lensing}
  \authorrunning{Diego et al.}

   \author{J. M. Diego
         \inst{1}\fnmsep\thanks{jdiego@ifca.unican.es}
       \and
         Sung Kei Li \inst{2} 
        \and 
        Alfred Amruth  \inst{2}
        \and 
        Ashish K. Meena \inst{3}
       \and
       Tom J. Broadhurst \inst{4,5,6}
       \and     
       Patrick L. Kelly \inst{7,8}
       \and
       Alexei V. Filippenko \inst{9}
       \and
       Liliya L. R. Williams \inst{7,8}
       \and
       Adi Zitrin \inst{3}
       \and
       William E. Harris \inst{10}
       \and
       Marta Reina-Campos \inst{10,11}
       \and
       Carlo Giocoli \inst{12,13}
       \and 
       Liang Dai \inst{14}
       \and
       Mitchell F. Struble\inst{15}
       \and 
       Tommaso Treu  \inst{16}
       \and
        Yoshinobu Fudamoto \inst{17}
       \and
       Daniel Gilman  \inst{18,19}
       \and
       Anton M. Koekemoer \inst{20}
       \and
       Jeremy Lim \inst{2}   
       \and
       J.M. Palencia \inst{1}
       \and
       Fengwu Sun \inst{21}
       \and
       Rogier A. Windhorst  \inst{22}
    }      
   \institute{Instituto de F\'isica de Cantabria (CSIC-UC). Avda. Los Castros s/n. 39005 Santander, Spain 
        \and
        Department of Physics, The University of Hong Kong, Pokfulam Road, Hong Kong  
        \and
         Physics Department, Ben-Gurion University of the Negev, P.O. Box 653, Be’er-Sheva 84105, Israel 
         \and        
          Department of Physics, University of the Basque Country UPV/EHU, E-48080 Bilbao, Spain 
         \and
          DIPC, Basque Country UPV/EHU, E-48080 San Sebastian, Spain 
          \and
          Ikerbasque, Basque Foundation for Science, E-48011 Bilbao, Spain 
         \and
         Minnesota Institute for Astrophysics, University of Minnesota, 116 Church Street SE, Minneapolis, MN 55455, USA 
         \and 
         School of Physics and Astronomy, University of Minnesota, 116 Church Street, Minneapolis, MN 55455, USA  
         \and 
        Department of Astronomy, University of California, Berkeley, CA 94720-3411, USA 
         \and
         Department of Physics \& Astronomy, McMaster University, 1280 Main Street West, Hamilton L8S 4M1, Canada 
         \and
         Canadian Institute for Theoretical Astrophysics (CITA), University of Toronto, 60 St. George St., Toronto M5S 3H8, Canada 
         \and
         INAF -- Astrophysics and Space Science Observatory of Bologna. Via Piero Gobetti 93/3, I-40129 Bologna, Italy 
         \and 
        INFN--Sezione di Bologna, Viale Berti Pichat 6/2, 40127 Bologna, Italy \label{2}  
         \and 
          Department of Physics, University of California, 366 Physics North MC 7300, Berkeley, CA 94720, USA       
         \and
         Department of Physics and Astronomy, University of Pennsylvania, 209 South 33rd Street, Philadelphia, PA 19104, USA
         \and
         Physics \& Astronomy Department, University of California, Los Angeles, CA, 90095, USA
         \and
         Center for Frontier Science, Chiba University, 1-33 Yayoi-cho, Inage-ku, Chiba 263-8522, Japan 
         \and 
         Department of Astronomy $\&$ Astrophysics, University of Chicago, Chicago, IL 60637, USA 
         \and
         Brinson Prize Fellow
         \and
         Space Telescope Science Institute, 3700 San Martin Drive, Baltimore, MD 21218, USA 
        \and 
        Steward Observatory, University of Arizona, 933 N. Cherry Ave., Tucson, AZ 85721, USA
        \and
         School of Earth and Space Exploration, Arizona State University, Tempe, AZ 85287-1404, USA 
          }

 \abstract{
Recent observations of caustic-crossing galaxies at redshift $0.7\lesssim z \lesssim 1$  show a wealth of transient events. Most of them are believed to be microlensing events of highly magnified stars. Earlier work predicted such events should be common near the critical curves (CCs) of galaxy clusters (``near region''), 
but some are found relatively far away from these CCs (``far region''). We consider the possibility that substructure on  milliarcsecond scales  (few parsecs in the lens plane) is boosting the microlensing signal in the far region. We study the combined magnification from the macrolens, millilenses, and microlenses (``3M-lensing"), when the macromodel magnification is relatively low (common in the far region). After considering realistic populations of millilenses and microlenses, we conclude that the enhanced microlensing rate around millilenses is not sufficient to explain the high fraction of observed events in the far region. 
Instead, we find  that the shape of the luminosity function (LF) of the lensed stars combined with the amount of substructure in the lens plane determines the number of mcirolensing events found near and far from the CC.   
By measuring $\beta$ (the exponent of the LF), and the number density of microlensing events at each location, one can create a pseudoimage of the underlying distribution of mass on small scales. We identify two regimes: (i) positive-imaging regime where $\beta>2$ and the number density of events is greater around substructures, and (ii) negative-imaging regime where  $\beta<2$ and the number density of microlensing events is reduced around  substructures. This technique opens a new window to map the distribution of dark-matter substructure down to $\sim 10^3\, \Msun$. 
We study the particular case of seven microlensing events found in the Flashlights program in the Dragon arc ($z=0.725$).
 A population of supergiant stars having a steep LF with $\beta=2.55^{+0.72}_{-0.56}$ fits the distribution of these events in the far and near regions. 
We also identify a small region of high density of microlensing events, and interpret it as evidence of a possible invisible substructure, for which we derive a mass of $\sim 1.3 \times 10^8\,\Msun$ (within its Einstein radius) in the galaxy cluster. 
   }

   \maketitle
%

\section{Introduction}

Galaxy clusters are the most powerful lenses in the universe. At the critical curves (CCs hereafter), and ignoring microlenses, small sources can be magnified by very large factors, with the maximum magnification for a source of radius $R$, $\mu_{\rm max} = \mu_o/\sqrt{R} $, where $\mu_o$ is a constant related to the smoothness of the lensing potential.
For galaxy clusters, $\mu_o$ can be of order 10 when $R$ is expressed in arcseconds. At the caustics of these clusters, stars with sizes a few times $\Rsun$ (that is, $R \approx 10^{-11}$ arcseconds at redshift $z \approx 1$) can reach theoretical extreme magnification factors exceeding $10^6$ \citep{Miralda1991}. In practice, the ubiquitous presence of microlenses from the intracluster medium (ICM) reduces the maximum magnification for these stars to $< 10^5$ \citep{Venumadhav2017,Diego2018}. Despite this reduction in the maximum magnification due to microlenses, the flux from massive lensed stars at $z \approx 1$ that are at a fraction of a parsec from a cluster caustic can be boosted by $\sim 7$--10 mag and be detected with current telescopes reaching a depth of 28 mag \citep{Kelly2018,Golubchik2023,DiegoBUFFALO}.

The extreme magnification near the critical curves of clusters has allowed the discovery of distant stars that would otherwise remain undetected. The first such star, Icarus at $z=1.49$, was discovered \citep{Kelly2018} with the {\it Hubble Space Telescope (HST)}, and was quickly followed by many others also observed with {\it HST} \citep{Rodney2018,Chen2019,Kaurov2019,DiegoGodzilla,Welch2022,Kelly2022,Meena2023a}. The farthest star discovered to date with {\it HST} through this technique is Earendel at a record breaking $z\approx 6$ \citep{Welch2022}
In total, {\it HST} has already discovered several dozen lensed-star candidates at $0.725 < z < 6$ \cite{Kelly2022}, most of them believed to be blue supergiants (BSGs) and luminous blue variable stars (LBVs). {\it HST} has passed the torch to the new {\it James Webb Telescope (JWST)}, which in a short time has already discovered over a dozen lensed-star candidates \citep{Chen2022,Diego2023Gordo,Meena2023b,Furtak2024,DiegoMothra,Yan2023}. Among these, several are believed to be red supergiants (RSGs), which are difficult to detect with {\it HST} \citep{Diego2023Gordo,DiegoMothra,DiegoBUFFALO,Yan2023}.  {\it JWST} will extend the search for distant stars to even higher redshifts and also to fainter stars. With a little luck, {\it JWST} will even directly observe the first generation of stars (Pop~III) in caustic crossing high-redshift galaxies \citep{Windhorst2018}.

Some of these transients are believed to be due not to microlensing, but to intrinsic variability of LBVs  that can increase their brightness by several magnitudes \citep{Weis2020}. They have up to 5\,mag variations on decade-long timescales, and smaller amplitudes on shorter timescales of months to years that are typical of supergiants. LBVs are luminous enough that they can be observed even at modest magnification factors ($\mu \approx 20$) if they are at $z<1$ (a star at $z=1$ with $L=10^6\, \Lsun$ would have apparent magnitude $\sim 28.7$ at $\mu=20$). Owing to their variable nature, they can be identified as transients in difference images between two epochs. 
At higher redshift, even these bright stars would become undetectable unless they are magnified by larger factors (a star at $z=2$ with $L=10^6\, \Lsun$ would have apparent magnitude $\sim 32.4$ at $\mu=20$ and $\sim 28.2$ if $\mu=1000$).  
Although most of the lensed stars are found in regions near cluster CCs, a significant fraction of these stars have been observed farther from the CCs where the magnification from the cluster is relatively small ($\mu < 100$). Examples include the off-caustic event described by \cite{Meena2023a} or some of the events reported by \cite{Kelly2022} and \cite{Yan2023}. 

The outbursts of LBVs  can be confused with genuine microlensing events, especially if the observations are separated by long periods that do not allow us to distinguish a microlensing event from an LBV outburst based on the light curve. A genuine microlensing event (in the optically thin regime) near the microcaustic (or maximum magnification) has a well-defined shape for the light curve since the luminosity changes as $1/\sqrt{(t-t_o)}$, where $t$ is time and $t_o$ is the time at which the background star touches the microcaustic ($t_o$ is a free parameter).  
LBVs are very rare compared with the more numerous but fainter supergiant stars, and since we can only identify them through their outbursts (or active phase), active LBVs  are even rarer, so we expect to see only a  few of them. The specific number depends on their abundance in the host galaxy, driven primarily by the recent star-formation history of that galaxy; hence, we expect to see them in very blue portions of lensed galaxies. Despite their scarcity, but because of their high luminosity and varying nature, LBVs  are good candidates for transient events that take place in regions of low magnification.

In the lens plane, the magnification at a short distance, $d$, from the CC can be well approximated by $\mu\approx \Theta('')/d('')$ \citep{Schneider1992}. 
In this expression $\Theta('')$ is related to the inverse of the derivative of the lensing potential at that position. For a symmetric lens, $\Theta('') = {\rm constant}$, and for an isothermal profile, it is exactly the Einstein radius, but for real nonsymmetric lenses with elliptically shaped CCs,  $\Theta$ varies along the CC, with maximum values at the cusps of caustics. For massive clusters where lensed stars have been discovered, $\Theta('')$ takes values between $\sim 50''$ and $\sim 100''$. Then, for these clusters, and at distances $d\gtrsim 1''$, the magnification from the cluster typically drops below 100. 
 In these regions of the lens plane with $\mu < 100$, the combined effect from the macrolens and the microlenses is often subcritical, $\mu\times\Sigma_{*} \lesssim \Sigma_{\rm crit}$, for typical values of the surface mass density of microlenses found near CCs, $\Sigma_{*} < 20\, \Msun$. Near the CC, even for small values of $\Sigma_{*}$ there is always a region around the CC in which microlensing supercriticality is achieved,  $\mu\times\Sigma_{*} > \Sigma_{\rm crit}$. In this region, the probability of microlensing events is expected to be maximum \citep{Diego2018,Palencia2023}. We refer to this portion of the lens plane as the ``near region.'' In contrast, outside this region and away from the CC, $d$ increases and $\mu$ decreases with  $\mu\times\Sigma_{*} < \Sigma_{\rm crit}$. Here we are in the microlensing subcritical portion of the lens plane where microlensing events are more rare. We refer to this portion of the lens plane as the ``far region''  corresponding to the regions inside and outside the corrugated network of small critical curves around the galaxy cluster CC \citep[see, for instance,][ for a description of this corrugated network]{Diego2018}.
 
 It is in principle difficult to explain the apparently high number of events found in the far region. This begs the question of whether a significant fraction of these events are active LBVs which can be more easily observed in the far region, or the cluster lens model is inaccurate on small scales, lacking substructures in the far region that can boost the magnification, and hence become supercritical around the substructures. Perturbations in the mass distribution on scales comparable to small satellites in the cluster (millilenses) can create pockets of relatively high magnification on angular scales of several milliarcseconds at distances of few arcseconds from the CCs.  These pockets of high magnification become islands of supercriticality where microlenses along the line of sight can now create more frequent microlensing events. 
 The combined lensing effect of a galaxy cluster scale lens, with its swarm of small satellites and the myriad of microlenses from the matter associated with the ICM, has not been studied previously in detail. We refer to this effect as 3M-lensing (macromodel lenses, millilenses, and microlenses), and it constitutes one of the foci of this work.

%
\begin{figure*} 
   \includegraphics[width=18.0cm]{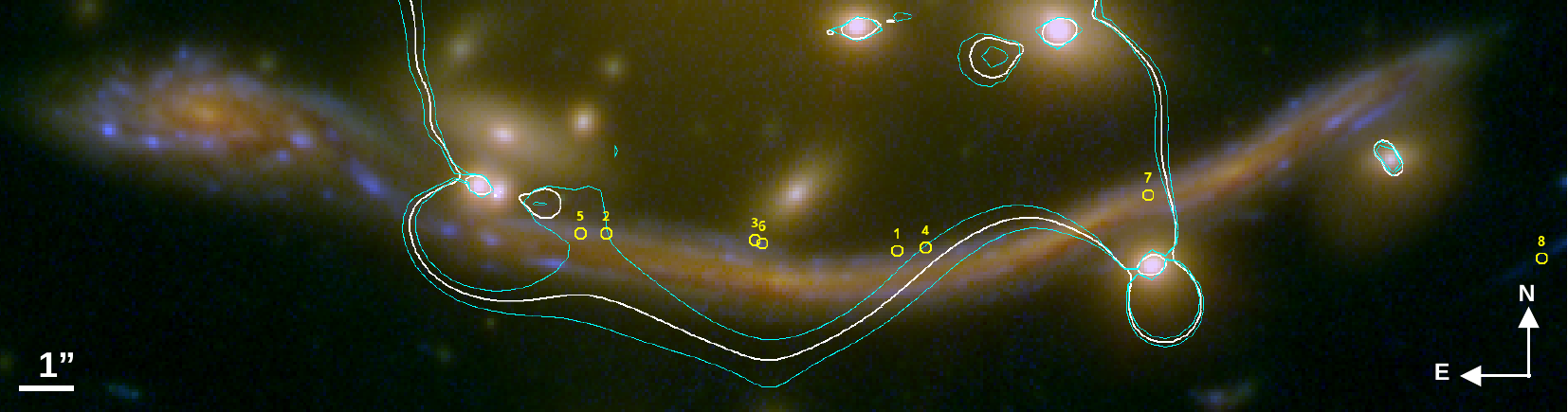}  
      \caption{
      Dragon arc as seen by {\it HST} (blue = F435W, green = F814W, red = F160W). The eight transients (seven in the arc) identified by \cite{Kelly2022} are marked with circles. Labels are the same as in the original reference. The white curve is the CC from our lens model (see Appendix) at the redshift of the arc. The two cyan curves mark the boundary region between macromodel magnification above and below 100. The arc covers $\sim 1150$\,kpc$^2$ in the lens plane. Out of this, 190\,kpc$^2$ is within the cyan curves (near region) and 960\,kpc$^2$ is  outside the cyan lines (far region). 
         }
         \label{Fig_Dragon}
\end{figure*}

Recent observations made with {\it JWST} of some of these cluster lenses have revealed a wealth of unresolved structures in the ICM \citep{Lee2022,Faisst2022,Harris2023}. Some of these objects are expected to be globular clusters (GCs) that are stripped away from their host galaxies by strong tidal forces from the cluster. These are the same forces that strip stars away from the infalling galaxies and into the ICM. An extended population of GCs in the ICM has been found, for example, in the rich lensing cluster Abell 2744 at $z=0.3$ with {\it JWST} imaging \citep{Harris2023}. In addition to GCs, the inner regions of galactic cores in small galaxies can survive tidal forces and appear as GC-like objects. These ultracompact dwarf galaxies (UCDs) tend to be more massive than GCs and possibly harbor a supermassive black hole (SMBH) in their center. For simplicity we refer from now on to all these unresolved objects as GCs,  keeping in mind that other types of objects may fall into this category. 

The number and distribution of these GCs is in agreement with observations made at lower redshift and also with expectations from numerical $N$-body simulations. Thousands of GCs with masses in the range $10^5\, \Msun < M <10^7\, \Msun$ are expected to be found within the critical curve of these clusters \citep{Faisst2022,Lee2022,Diego2023MACS0416,Harris2023}. These GCs can act as millilenses, whose lensing effect is magnified by the macrolens \citep{Gilman17,Dai2018,Williams2024}. In the vicinity of GCs, pockets of high magnification are created which, combined with the ubiquitous microlenses, can result in an increased rate of microlensing events around the millilenses, and near their small CCs around them, typically spanning a few milliarcseconds in the image plane. That is, for the smallest millilenses the increased rate of events would appear to originate from the same {\it HST} or {\it JWST} pixel (30 milliarcseconds for NIRCam short-wavelength detectors). 
Small dark matter (DM) structures can also act as millilenses, since these are predicted by many DM models \citep{Kolb1993,Graham2016,Visinelli2018,Arvanitaki2020,Gilman21,Gorghetto2022}. 
Microlenses overlapping with these small-scale DM structures make transient events more likely around them, serving as signposts of small-scale fluctuations in the distribution of DM. This is discussed in detail in this work.

The small CCs around these millilenses (or in general small DM structures) will inevitably overlap with the microlenses from the same ICM. The net lensing effect is a combination of the macrolens, the millilens, and the numerous microlenses. The effect of large macromodel magnifications plus microlenses has been studied in detail in earlier work \citep{Venumadhav2017,Diego2018,DiegoExtreme,Palencia2023}. The combined effect of large macromodel magnification plus millilenses was studied over two decades ago by (for example) \cite{Mao1998} and \cite{Metcalf2001}, and more recently by many others \citep[e.g.,][]{Hezaveh16,Gilman17,Gilman18,Dai2018,Cyr-Racine19,Gilman19,Gilman20a,Powell23,Gilman24,Williams2024,Tsang24}. The combination of the three effects has not been considered in detail so far, and to the best of our knowledge is presented here for the first time. 

In this work we pursue four goals: 
(i) study the macro+milli+micro lensing (``3M-lensing'') effect over 
stars at cosmological distances and near cluster CCs in order to provide context for recent and future discoveries of lensed stars where 3M-lensing is likely taking place, 
(ii) address the question of whether the millilensing effect from the numerous millilenses is sufficient to explain the transient events found at distances $d>1''$ from cluster CCs, 
(iii) study the relation between the number of observed microlensing events, the amount of substructure on small scales in the lens plane, and the luminosity function of the background population of high-redshift stars, and 
(iv) apply our results to recent observations, in particular to the case of the seven alleged microlensing events found by {\it HST} in the Dragon galaxy at $z=0.725$ as part of the Flashlights program \citep{Kelly2022}. This arc was originally  known as the Giant Arc or A370 Arc01 \citep{Soucail1987,Soucail1988,Lynds1989,Grossman1989,Smail1993,Smail1996}, and rebranded as the Dragon arc after new images were obtained following the {\it HST} Servicing Mission 4 update of the ACS in 2009\footnote{https://www.newscientist.com/article/dn17765-upgraded-hubble-telescope-spies-cosmic-dragon/}.

The paper is organized as follows. 
Section~\ref{Sect_Def}
presents a series of definitions that are used throughout and gives examples of typical scales appearing in lensing that become useful in later portions of the paper. 
The simulations of the 3M-lensing effect used in this work are presented in Section~\ref{Sect_Sims} . 
We focus in Section~\ref{Sect_Stat1} on the probability of magnification in 3M-lensing.  
Section~\ref{Sect_Scaling} discusses the scaling of the effect with millilens mass and macromodel magnification. 
In Section~\ref{Sect_Stat2} we describe how to compute the contribution, from a given mass function of GCs, to the area in the source plane (which can be interpreted as a probability) where microlensing effects are expected to be maximum. 
Section~\ref{Sect_Results1} estimates the probability of microlensing events in the far region around millilenses, while 
Section~\ref{Sect_Results2} estimates the probability of microlensing events anywhere in the far region, not just near millilenses. 
In Section~\ref{Sect_MappingDM} we discuss how to apply the previous results to map the distribution of DM on small scales, and apply our results to the particular case of the Flashlight microlensing events in the Dragon arc. 
The Dragon arc has been observed by the VLT/MUSE, providing resolved spectral information along the arc \citep{Patricio2018}.
We discuss our results in Section~\ref{Sect_Discussion} and conclude in Section~\ref{Sect_Conclusions}. An Appendix contains details of the lens model for the particular example used to illustrate this work. 


\section{Definitions and useful numbers}
\label{Sect_Def}
We use several definitions throughout the paper, which for convenience we summarize here. 
Critical curves (CCs) are the regions in the image plane (also known as lens plane or plane of the sky) where magnification formally diverges. The image plane and observer (or source) plane are connected through the lens equation, $\beta=\theta-\alpha( \theta,M)$, where $\beta$ are positions in the source plane, $\theta$ are positions in the image plane and $\alpha(\theta,M)$ is the deflection angle that depends on the distribution of mass of the lens. Through this equation, we can map the CCs into the corresponding curves in the source (or observer) frame, which are called caustics. A caustic region is the portion of the source plane which is bounded by the caustic curves. 
The near region is defined as the portion of the lens plane close to the cluster CC where the rate of microlensing events is maximized. This region is defined in terms of the cluster magnification and the surface mass density of microlenses. It is  a band around the cluster CC where the cluster magnification is above the critical value, $\mu \gtrsim \mu_{\rm crit} = \Sigma_{\rm crit}/\Sigma_{*}$, where $ \Sigma_{\rm crit}$ and $\Sigma_{*}$ are the critical surface mass density for lensing and the surface mass density of microlenses, respectively \citep{Diego2018}. An example is shown in Figure~\ref{Fig_Dragon}, where the near region is contained within the two thin cyan curves. Similarly, the far region is the portion of the lens plane where the macromodel magnification is $\mu < \mu_{\rm crit}$, and in the same figure it would be the region outside the band defined by the two cyan curves. 

Following standard practice \citep[e.g.,][]{Treu10}, the term macrolens is used when referring to the galaxy cluster scale lens, and the term millilens is used when referring to GCs or in general unresolved structures such as galactic core remnants, dwarf galaxies in the ICM, satellites in general, small DM halos, or intermediate-mass primordial black holes \citep{Dike23}. These systems are expected to have Einstein radii of order milliarcseconds, hence the term millilensing. The term microlens is used for stars or stellar remnants in the ICM, which have Einstein radii of order microarcseconds.  Some DM candidates such as primordial black holes with masses comparable to stellar objects would also fit in this category \citep[see, for instance,][]{Diego2018,Oguri2018,Muller2024}. 

 For example, the Einstein radius of a 1 $\Msun$ microlens at $z=0.375$  and for a source at $z=0.725$ (the redshifts of the cluster lens and Dragon galaxy, respectively) is 1.8 microarcseconds ($\mu$as) before accounting for the effect of the macrolens or millilens. For the same redshifts, a millilens with mass $10^5\, \Msun$ would have an Einstein radius of 0.57 milliarcseconds (mas), also before accounting for macromodel effects. For any other mass, $M$, at the same redshift, the Einstein radius would be $\theta_E \approx 1\, {\rm mas} \times \sqrt{M/(3.1\times10^5\, \Msun)}$. In general, when embedded in a macromodel potential with magnification $\mu$, the CC around the millilens or microlens with mass $M$ behaves as a larger millilens or microlens with effective mass $\mu_t\times M$ \citep{Diego2018,Oguri2018}, where $\mu_t$ is the tangential macromodel magnification ($\mu_r$ would be the radial component and $\mu=\mu_t\mu_r$). For the particular case of a microlens near a millilens, the same scaling with magnification applies, only in this case the magnification $\mu_t$ is from the combined effect of the macromodel plus the millilens. 

The CCs associated with these types of lenses are macro-CCs, milli-CCs, and micro-CCs. Similarly, we use the terms macrocaustic, millicaustic, and microcaustic when referring to the corresponding caustics. We refer to the macromodel magnification as $\mu_{\rm 1m}$, while we use the term $\mu_{\rm 2m}$ when referring to the magnification from the combined macromodel plus millilens, and $\mu_{\rm 3m}$ (the 3M-lensing magnification) when referring to the magnification of all three components (macrolens plus millilens plus microlenses). 

In Section~\ref{Sect_MappingDM} we define the luminosity function (LF) of stars as $dN/dL = \phi(L)\propto (1/L)^{\beta}$, which gives the number of stars per luminosity bin and unit area.
This ``classic definition" is useful when working with nonmagnified and uniform distributions (or sources in the source plane before magnification is applied), since in this case the properties of the LF are independent of the region being considered. But when dealing with lensed sources, there is a strong dependence on the magnification. Because of this, we also use a different definition for the lensed luminosity function, or $\hat{\phi}(L)$, which gives the number of stars per luminosity bin and in a given area (not per unit area). This alternative definition is useful when we are considering the number of stars in a particular region with macromodel magnification $\mu$, in the interval $\mu_{\rm min} < \mu < \mu_{\rm max}$. That is, in this case $\hat{\phi}(L)$ means $\hat{\phi}(L,\mu_{\rm min}^{\rm max})$, where $\mu_{\rm min}^{\rm max}$ is all macromodel magnifications in the interval of magnification, but for convenience we simply use the expression $\hat{\phi}(L)$. 

Below we present a few useful numbers for the particular case of the Dragon arc, which holds the record for the number of transients discovered as part of the Flashlights program \cite{Kelly2022}. The location of these events in relation to the CC is shown in Figure~\ref{Fig_Dragon}. This arc contains seven high-significance transients, with at least two of them found in the far region (see Figure~\ref{Fig_Dragon}) and good candidates to be stars impacted by 3M-lensing \citep{Kelly2022}. 
For the particular case of the Dragon galaxy, the redshift of the lens is 0.375 (A370 cluster), and the redshift of the lensed galaxy is 0.725. We adopt a flat-universe cosmology with $\Omega_m=0.3$ and $h=0.7$. For this model, the angular diameter distances to the lens at $z=0.375$, the source at $z=0.725$, and from the lens to the source are 1066\,Mpc, 1495\,Mpc, and 646\,Mpc, respectively. For the same cosmology, $1''$ subtends 5.16\,kpc at $z=0.375$ and 7.24\,kpc at $z=0.725$. The critical surface mass density for these redshifts is $\Sigma_c=3640\, \Msun\,{\rm pc}^{-2}$ and the distance modulus to $z=0.725$ is 43.24\,mag. For illustration purposes, a star with absolute magnitude $-7$ (color corrected in a given filter) and magnified by a factor of 100 would have apparent magnitude 31.2, still out of reach of {\it JWST} with 1\,hr integration in one of the wide filters. However, the same star during a microlensing event (lasting typically a few days to a few weeks depending on the mass of the microlens, relative speed, and direction of motion with respect to the microcaustic), can be temporarily magnified by a factor of  $\sim 1000$,  and would appear $\sim 2.5$\,mag brighter (i.e., $\sim 28.7$\,mag) during that period. This would be detectable in a 1\,hr integration time with {\it JWST} and be interpreted as a transient. 

 Finally, we use the term ``detectable through microlensing" (or DTM) stars to refer to all the stars that have detectable changes in brightness due to microlensing. These are either stars that (i) are detected in several epochs but between two epochs change their brightness by some amount (due to a microlensing event), or (ii) are   detected in only one epoch because  microlensing is temporarily boosting their flux. In general, we assume that the second type of DTM stars are $\sim 2$\,mag below the detection threshold before microlensing. During a microlensing event, any DTM star increases (or decreases) its brightness by $\sim 2$\,mag and can be recognized as  a transient.

\section{3M-lensing}
\label{Sect_Sims}

To study the 3M-lensing effect, we rely on simulations that combine all three mass ranges (macro, milli, and micro). Since our focus is to study microlensing events around millilenses (embedded in a macrolens potential), we set the simulation parameters to match the scale of millilenses but at the same time resolve the microlenses. As mentioned in Section~\ref{Sect_Def}, before macromodel effects, the scale of a $10^5\, \Msun$ millilens is typically $\sim 1$\,mas in the image plane. On that scale, the effect of the macromodel can be very well approximated as a smooth gradient with a slope that 
is roughly the inverse of the Einstein radius of the macrolens \citep{Diego2018}. 
The scale of microcaustics is $\sim 1$\,$\mu$as. To properly resolve microcaustics, the pixel size needs to be much smaller than 1\,$\mu$as. A pixel size of 10 nanoarcseconds (nas) in the source plane is sufficient to resolve the microcaustics from the smallest microlenses. After accounting for macromodel effects, the critical region of the millilens grows as the macromodel magnification. Hence, the simulation needs to span several milliarcseconds in the image plane if the macromodel magnification is $\mu_{\rm 1m}>10$. To simulate several milliarcseconds in the image plane with a resolution of 10\,nas in the source plane would require a prohibitive number of $\sim 10^{12}$ pixels. We can significantly reduce this by simulating a smaller millilens, since the number of pixels needed scales approximately with the mass of the millilens. Luckily, the magnification properties of 3M-lensing for larger millilenses can be extrapolated by simply rescaling the results derived with smaller millilenses (see Section~\ref{Sect_Scaling}). In particular, we consider very small millilenses with masses of order $10^3\, \Msun$ and later study how our results scale with the millilens mass.

\begin{figure} 
   \includegraphics[width=8.6cm]{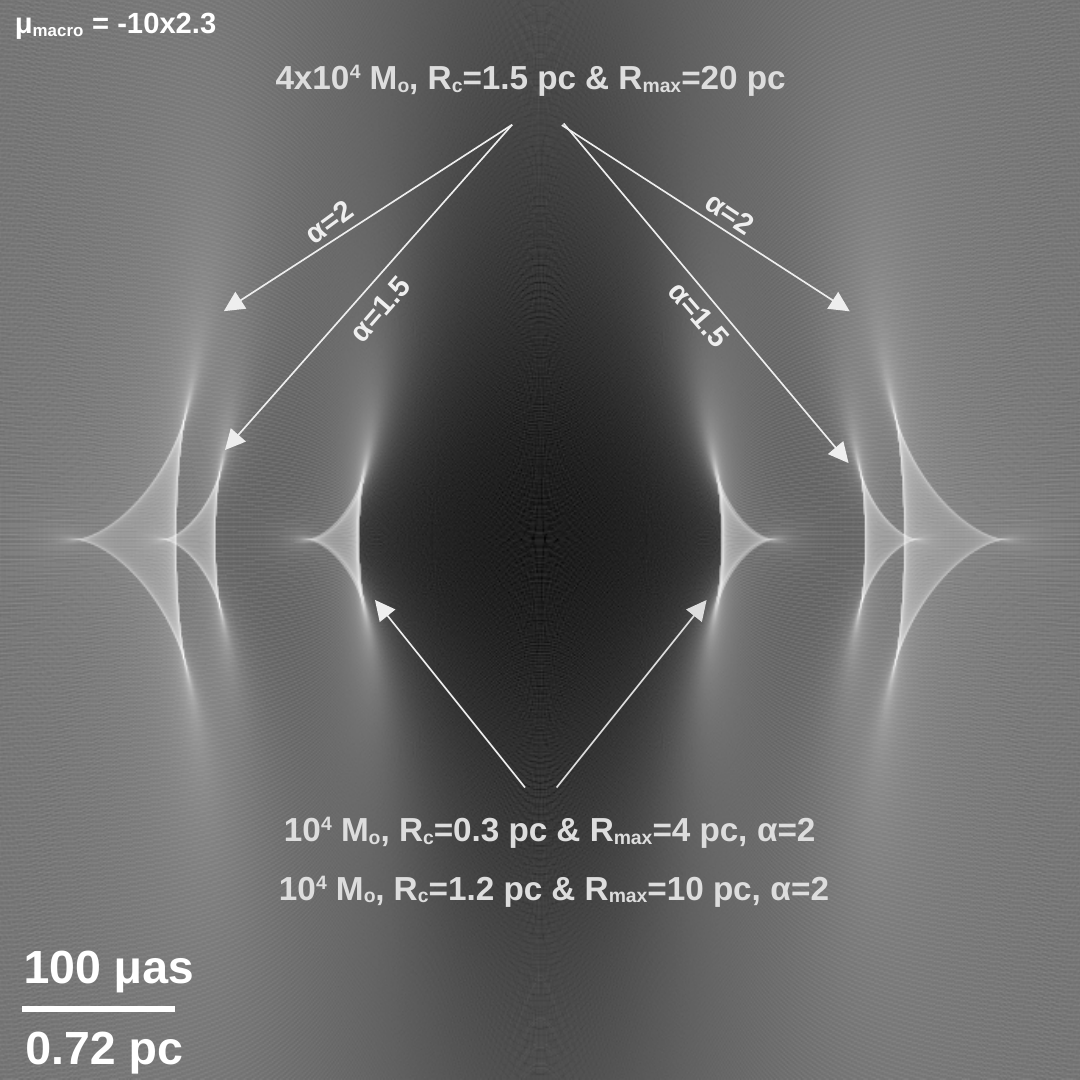}
      \caption{Effect of the millilens profile. Comparison of millicaustics for four millilenses under the influence of the same macromodel magnification ($|\mu_{\rm 1m}|=23$) but for different mass, core size $R_c$, truncation radius $R_{\rm max}$, and exponent $\alpha$. The profile is defined as $\rho(r) \propto (R_c +r)^{-\alpha}$. The image shown in grayscale is the 
      sum of the four magnifications from the four millilenses. The caustics for the two millilenses with mass $10^4\, \Msun$ and slope $\alpha=2$ are nearly identical and fall on top of each other, indicating that the mass is the main driver defining the size of the caustic region. The  largest millicaustic corresponds to a millilens  with 4 times more mass, and larger core and truncation radii, but the same slope $\alpha$. The area above $\mu=100$ is a factor of 4  larger than in the smaller millilenses. A third millilens with the same mass, $R_c$, and $R_{\rm max}$ but a shallower profile ($\alpha=1.5$) behaves as the larger millilens with $\alpha=2$ but a mass of $2.93\times10^4\, \Msun$, owing to the reduction in mass within the Einstein radius. Even shallower profiles ($\alpha \lesssim 1$) with large cores result in subcritical millilenses (no caustics or cusps). On the other hand, a steeper profile with $\alpha=3$ or greater produces a millicaustic almost indistinguishable from the one obtained when $\alpha=2$.
         }
         \label{Fig_Millicaust_Comp}
\end{figure}
%

Hence, to explore 3M-lensing in a wide range of scenarios, we define a fiducial model that is used for the main calculations and later study the scaling with macromodel, millilens, and microlenses around this fiducial model. For the fiducial model we adopt a macromodel magnification $\mu_m=\mu_t\times\mu_r=\pm 10\times2.3$, where $\mu_t$ and $\mu_r$ are respectively the tangential and radial magnifications from the macromodel. The magnification can be positive or negative depending on what side of the critical curve we are considering. The side with positive magnification is also the side with positive parity (counterimages have the same orientation as the original source). In contrast, when the magnification is negative the counterimage has negative parity (inverted in relation to the original source). The  assumed macromodel magnification is small enough such that the microcaustics do not usually overlap and microlensing is a rare event. For the millilens we assume a fiducial model with a relatively small GC having a mass of $2\times10^3 \, \Msun$, with 
a truncated core power-law density profile
\begin{equation}
\rho(r)_{3D} \propto \frac{1}{(R_c+r)^{\alpha}}\, , 
\label{eq_rhoalpha}
\end{equation}
where $R_c$ is the core radius and the profile is truncated at some radius $R_{\rm max}$.
This millilens is representative of a small and compact GC that would survive the strong tidal forces in clusters. 
The core radii of the density profiles of millilenses can be approximately estimated from dwarf galaxies. Typical radii of $10^9\, {\rm M}_\odot$ galaxies in the LITTLE THINGS galaxy survey are about 300\,pc, with substantial variation between individual galaxies \cite[Table 2 of][]{Oh2015}. This size is consistent with those presented by \cite{Wolf2010}. 
Motivated by the virial condition and assuming that the concentration parameter is the same for all subhalos, $r_{\rm core}\propto m^{1/3}$. This relation can be scaled to smaller masses \citep{Williams2024}. 
For instance, for  very small millilenses with mass $2\times10^3\, \Msun$, the core radii should be a factor of $\sim 80$  smaller than for the $10^9\, \Msun$ halo, or $\sim 3.7$\,pc. 
For our calculations we adopt the most optimistic scenario where millilenses are most lensing-efficient, and therefore we assume a much smaller core radius of $r_c=0.15$\,pc.  For the truncation radius we take $\sim 10$ times the core radius. Very compact structures in the Milky Way, such as the central region of R136 in the Large Magellanic Cloud (LMC), would have a similar scale \citep[diameter $\approx 1$\,pc from][]{Massey1998}. Our core and truncation radii for the small  $2\times10^3\, \Msun$ GC are  also consistent (after extrapolation to smaller masses)  with the radii of the more massive GCs found in the Milky Way by \cite{Baumgardt2018}, who find typical  half-mass radii of $\sim 5$\,pc for GCs with mass $\sim 10^5\, \Msun$. 
A small core radius also accounts for the fact that we expect the more compact structures to be the ones surviving in denser environments \citep{Moline2017}. 

\begin{figure*} 
   \includegraphics[width=18.2cm]{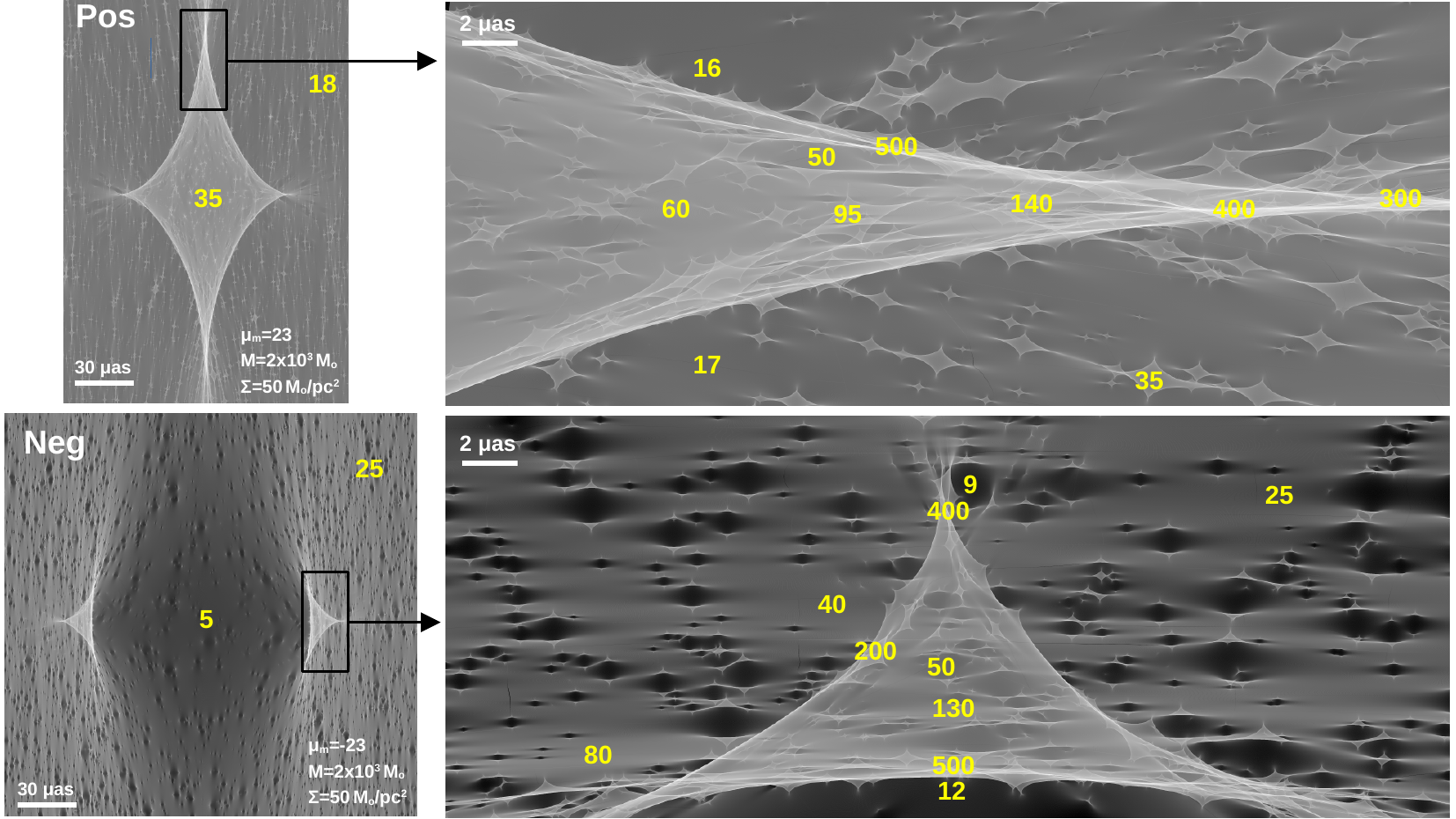}
      \caption{Simulated magnification maps of 3M-lensing. {\it Left panels.} Shown as grayscale is the log of the magnification in the observer plane (caustics) around a millilens with mass $2\times10^3 \, \Msun$, in two regions (positive and negative parities) where the macromodel magnification is $\pm 23$ and with a surface mass density of microlenses $\Sigma=50\, \Msun {\rm pc}^{-2}$. The numbers in yellow indicate typical magnification values at these positions. For millilenses in regions with positive parity, outside the millicaustic region the magnification is typically below the macromodel value (but higher at the microcaustic regions). In the center of the millicaustic region, the typical magnification is $\sim 50\%$ higher than the macromodel value. This situation is reversed in the region with negative parity. 
      {\it Right panels.}  Zoom-in around the regions of highest magnification at the millicaustics and marked with black rectangles in the left panel. Near the millicusps and millicaustics, microcaustics always overlap one another at magnifications greater than 100, hence maximizing the occurrence of microlensing events. 
         }
         \label{Fig_3M}
\end{figure*}

The specific shape of the profile and truncation radius play a role in the lensing effect since they define the mass contained within the Einstein radius of the millilens. As mentioned above, in this work we consider the most favorable condition where the millilenses are very compact  and most of their mass is contained within the Einstein radius. This is satisfied when $\alpha=2$ or greater. In this situation, the dependence on the profile is very weak. Only for shallow profiles ($\alpha \lesssim 1.3$) and large cores, the millilens may be subcritical and not able to produce large magnification factors.  
A visual comparison of the millicaustics for four different millilens models is shown in Figure~\ref{Fig_Millicaust_Comp}. The macromodel magnification for the four millilenses in the figure is set $\mu_{\rm 1m}=\mu_t\times\mu_r=-10\times2.3=-23$. The two smallest millilenses have exactly the same mass and produce millicaustics that are nearly identical, despite the two millilenses having different core sizes and truncation radii (but the same $\alpha$). The two millilenses with larger mass have a correspondingly larger millicaustic area. For the large millilens with $\alpha=2$, the gap between the caustic regions (demagnification region) increases as the square root of the mass when compared to the smaller millilenses with the same $\alpha$, so a millilens 100 times more massive can demagnify a region 10 times larger in diameter. For the millilens with identical mass but a shallower profile ($\alpha=1.5$), we observe a reduction in the lensing probability (or area with magnification greater than some value) of $\sim 25\%$. 
On the other hand, a steeper slope with $\alpha=3$ \citep[and consistent with $N$-body simulations of subhalos;][]{Moline2017} 
increases the lensing probability, but only by $\sim 2\%$, so our choice of $\alpha=2$ is valid to represent even more compact millilenses with $\alpha>2$. 


Finally, to complete our fiducial model for the 3M-lensing simulations, for the microlenses we consider a surface number density of $\Sigma_{*} = 50\,\Msun\,{\rm pc}^{-2}$.  This 
is close to the expected value around the Dragon arc, if one assumes that stars in the ICM contribute $\sim 2\%$ to the total projected mass at this position. The value is also consistent with direct estimates of the surface mass density of stars in the intracluster light (ICL) from recent {\it JWST} data in massive clusters and at distances between 50\,kpc and 70\,kpc from the center of the cluster \citep{Montes2022},  the distance at which our case study (the Dragon arc) is from the center of A370.

Originally, the pixel scale is set to 30\,nas, and in the lens plane we distribute the microlenses randomly in a circular region of radius 1.2\,mas until we reach the desired surface mass density of $\Sigma_{*} = 50\,\Msun\,{\rm pc}^{-2}$. For the mass function of the microlenses we adopt a \citet{Chabrier2003} model with a lower mass of 0.1\,$\Msun$. The specific model for the mass function plays a secondary role, since the most relevant parameter for microlensing is the value of $\Sigma_{*}$. The considered circular area is sufficiently large to easily accommodate the small millilens of the fiducial model. A second higher resolution simulation is later done around the cusps of the millicaustics and with a smaller pixel size of 10\,nas that resolves the microcaustics even better. 

Since the macromodel magnification can be positive \citep[outside the cluster CC or positive-parity region;][]{Blandford1986} or negative (interior to the cluster CC or negative-parity region), and the millilens caustics behave very differently depending on the parity, we simulate both parities but keep the absolute value of the macromodel magnification constant. When simulating the two parities, we only change the tangential component of the macromodel magnification --- that is, we take the two values $\mu_t=\pm 10$. In tangential critical curves, the tangential magnification changes rapidly as one gets closer to (or farther away from) the CC, while the radial component of the macromodel magnification changes very slowly. A value of $\mu_t = \pm 10$ is representative of scenarios similar to the far region, where macro+microlensing alone is unlikely to produce transient events but the combined 3M-lensing effect can boost the probability of transients around millilenses in the far region.

The magnification in the observer plane (caustics) is computed using standard ray tracing.  We show the result for the fiducial model in Figure~\ref{Fig_3M}, and for the two parities, that is for $\mu_t=10$ (positive parity) and $\mu_t=-10$ (negative parity). The radial magnification is identical in both cases, $\mu_r=2.3$. The left panels show the caustic region with the 30\,nas pixel size while the right panels display the higher resolution simulation with 10\,nas per pixel and around two of the cusps of the millilens caustics. In all cases, the magnification (grayscale) is shown in log scale to better appreciate the details. The numbers in yellow indicate the typical magnification (from the macrolens and millilens) outside the caustic region and near the center of the caustic region. 

At the caustics the magnification can be very large. For these simulations the maximum magnification is limited by the nonzero size of the pixel but still results in magnification factors of $\sim 1000$ at the caustics for the 30\,nas pixel and a few thousand for the 10\,nas pixel. A large star at $z=0.725$ with $R \approx 100\, {\rm R}_{\odot}$ would be $\sim 33$ times smaller than this pixel size, and the maximum magnification at the caustic would be $\sim 6$ times larger. 

The case with positive parity (top-left panel) shows the classic diamond-shaped caustic. In the simulations, the larger tangential magnification from the macromodel goes in the horizontal direction, resulting in a caustic that is more stretched in the vertical direction. The magnification near the center of the caustic is almost twice the magnification of the macromodel, so most of the inner-caustic region provides a relatively modest boost in relation to the macromodel value. Only in the small regions near the four cusps of the caustic, and very close to the caustics themselves, the magnification from the millilens alone can be sufficiently large to make luminous stars at $z=0.725$ detectable. Immediately outside the caustic region the most common value for the magnification is below the macromodel value. In this outer region the effect of the millilens is to slightly demagnify sources, hence compensating the larger magnification inside the millicaustic region, and ensuring that the average magnification over sufficiently large areas equals the macromodel value (flux conservation). A source which is significantly larger than the millilens caustic, for instance a star-forming region several parsecs in size,  will have an average magnification very close to the macromodel value and thus insensitive to the presence of the millilens. Only very small objects within such a source, for instance stars,  can attain large magnification values when they are near the millicusps or millicaustics. 

For the case of negative parity (bottom-left panel) we observe some significant differences, with two small triangle-shaped high-magnification regions bracketing a larger low-magnification region. This is a well-known configuration for caustics in negative-parity regions \citep{Chang1979,Chang1984}. The magnification between the two triangular-shaped caustic regions can be very small, of order 1. A small object with a size of 1\,pc or less placed in this inner region would be demagnified by the millilens, making its detection more difficult. 
This scale would be larger for heavier millilenses or larger macromodel magnification values. Hence, it is possible that sources a few pc in size such as GCs or small star-forming regions in the lensed galaxy get demagnified by a millilens and remain undetected if their lensed counterimage is in a negative-parity region behind a millilens. This cannot happen for counterimages behind the millilens in the portion of the lens plane with positive parity, where demagnification more than a few percent cannot take place. Since sources near a cluster caustic form two highly magnified counterimages near the CC, one counterimage with positive parity and one counterimage with negative parity, objects as small as a star or a small group of stars may appear highly magnified on one side of the CC (positive parity) and remain undetected on the other side of the CC (negative parity). This mechanism could explain the lack of asymmetry between the positive- and negative-parity images of stars or groups of stars recently observed in highly magnified galaxies \citep{DiegoMothra,DiegoGodzilla,AngelaNature2024}. On smaller scales, a similar mechanism but involving microlenses was used to explain the lack of counterimages of lensed stars such as Icarus \citep{Kelly2018}. 

At the smaller microlens level, we show in the right panels a zoomed-in version of the high magnification near a diamond-shaped cusp (positive parity) and a triangular-shaped caustic region (negative parity). In both cases we see how microcaustics adopt similar shapes (diamonds and triangles) and have a tendency to align with the millicaustics. In some cases, microlenses around the millilens on the side with positive parity behave as microlenses with negative parity and vice versa for the side with negative parity. These rare exceptions can be appreciated near the cusp regions, where locally the parity can be inverted owing to the influence of the millilens. As expected, the number density of microcaustics increases in the cusps and near the caustics. This is due to the larger magnification of the millilens that concentrates more microcaustics in these regions. The size of the microcaustics also grows with the millilens magnification in a fashion similar to that near the caustics of galaxy clusters. The highest probability of observing microcaustic crossings is then near the cusps of millilens caustics. As described in earlier work \citep{Diego2018,Palencia2023}, when the effective surface mass density of microlenses approaches the critical value, $\Sigma_{\rm crit}$, microlensing effects are maximized (in particular, fluctuations in the observed flux). For our fiducial model, this happens when the combined magnification from the macrolens and millilens is $\mu_{\rm 2m} \gtrsim \Sigma_{\rm crit}/\Sigma_{*}\approx 75$.  
The right panel of Figure~\ref{Fig_3M} shows this effect near the cusps. For convenience, the baseline magnification (that is, $\mu_{\rm 2m}$) is marked at different positions. In the case of millilens cusps in positive-parity regions (top-right panel), the magnification just outside the cusp is typically 25\% smaller than the macromodel value. Inside the cusp region the magnification is higher than the macromodel and can exceed values of $\mu=100$ near the cusps and caustics. There are also small areas around a few microcaustics where the parity is inverted and the magnification can be relatively smaller. One such example is marked with the magnification value 50 in the top-right panel.

For millilenses in regions with negative macromodel parity (bottom-right panel), the most striking difference is the regions with significant demagnification. Outside the caustic region (or triangle), microcaustics can demagnify (with respect to the macromodel) regions as big as $R=0.01$\,pc, for instance the optical portion of a quasar accretion disk or a supernova photosphere months after the explosion\footnote{This area would be even larger in regions with higher macromodel magnification.}. In regions not containing a microcaustic, outside the caustic region the typical magnification ranges between $\mu\approx 25$ and $\mu\approx 40$ --- that is, between $\sim 10\%$ and $\sim 75\%$ higher than the macromodel value --- again compensating the lower magnification between the two caustic regions. Inside the caustic region the typical magnification is higher, especially near the cusps of the caustics. There is a sharp transition between the main caustic in the bottom portion of the figure where the magnification changes rapidly between extreme values to values of order 10. Inside the caustic region we also observe local changes in the parity, for instance around the microlens marked with magnification 200. As in the previous examples, near the cusps the microcaustics overlap, filling the space, and the probability of microlensing is maximum. In both examples we see this effect when $\mu_{\rm 2m} \gtrsim 100$, close to the value $\mu_{\rm 2m} \approx 75$ derived above. We adopt this value ($\mu_{\rm 2m} = 100$) as the critical magnification above which microcaustics are constantly overlapping and microlensing effects are maximum --- that is, in what follows we assume $\mu_{\rm crit}=100$. In Section~\ref{Sect_Discussion} we discuss how our results depend on this choice.


\section{Statistics of 3M-lensing magnification near a single millilens}
\label{Sect_Stat1}

\begin{figure} 
   \includegraphics[width=8.6cm]{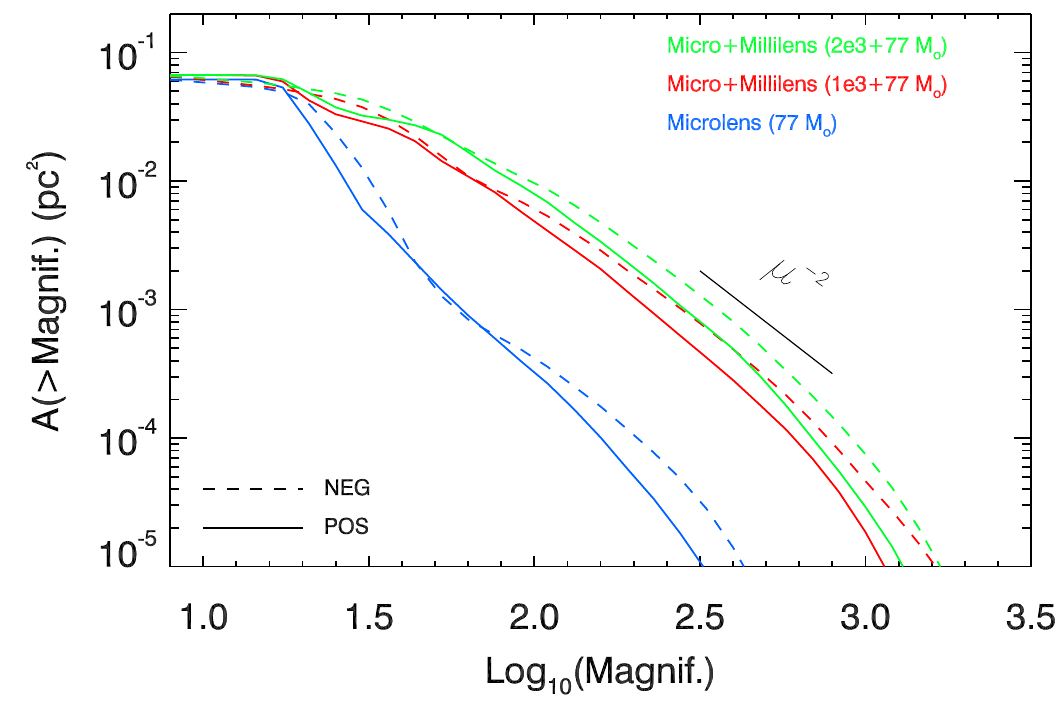}
      \caption{Probability of magnification in 3M-lensing. Blue lines are for macromodel plus microlenses only, while red and green lines are for 3M-lensing and for two millilens masses. Dashed lines correspond to negative parity and solid lines to positive parity. The probability scales as the total small-scale mass (millilens plus microlenses). 
         }
         \label{Fig_Histo1}
\end{figure}

The magnification pattern discussed in the previous section is interesting to interpret some events, and in particular to explain the lack of symmetry between pairs of images close to CCs when small objects, up to a few pc in size, are multiply imaged. In this work we are interested in the regime where the macromodel magnification is not that large, farther away from the CC, and in particular on the probability of having microlensing events around millilenses. For this it is useful to compute the area in the source plane having magnification above $\mu_{\rm crit}$, or $A(\mu>\mu_{\rm crit}$, since stars in this area are the most likely to show microlensing effects. 



We compute  $A(\mu>\mu_{\rm crit}$ in the two regions shown on the right side of Figure~\ref{Fig_3M},  for the two parities and around the cusps  of the millilenses. We compare with the area computed in the same region and for the same configuration of microlenses but removing the millilens. The result is shown in Figure~\ref{Fig_Histo1}. Dashed lines refer to the area computed in portions of the lens plane with negative parity (bottom-right panel of Figure~\ref{Fig_3M}), while solid lines are for positive parity (top-right panel of Figure~\ref{Fig_3M}). The green lines are for a millilens with mass $2\times10^3\, \Msun$ plus microlenses, while the blue curves are for the case where only microlenses are included in the simulation. For comparison, we show as red lines the case where the mass of the millilens is reduced by a factor of 2. 
As in the case of microlensing near caustics explored in earlier work, the probability of high magnification is slightly larger in areas with negative parity (dashed lines). In these regions significant demagnification can take place in relatively large areas, that is compensated by the larger magnifications of the cusps.  

In all cases, the probability of magnification scales as the expected $\mu^{-2}$ power law. The departure from this scaling at $\mu>1000$ is mostly an artifact due to the nonzero pixel size, although at larger magnification factors of $\mu> 10,000$ many microcaustics overlap and the magnification is expected to fall faster than $\mu^{-2}$ and become a log-normal distribution \citep{DiegoExtreme,Palencia2023}. 
The ratio of the green to the blue curves corresponds approximately to the ratio of masses between the millilens and the stellar mass in the same region. For this particular area the stellar mass in the right panels of the figure is roughly the fiducial value times the area of the two right panels and times the macromodel magnification (to transform the source area into image area): $M_{*}=(50\, \Msun\, {\rm pc}^{-2})\times (0.163\, {\rm pc})\times(0.41\, {\rm pc})\times 23 = 77\, \Msun$. Dividing the millilens mass ($2\times10^3\, \Msun$) by this mass gives a ratio of 27, which is roughly the ratio between the green and blue lines. Similarly, reducing the mass of the millilens by a factor of 2 results in a reduction in the probability by approximately the same factor (red curves). 


Although not shown in the figure, the corresponding probability for the case where microlenses are ignored would be very similar to the fiducial model but a bit below the green lines owing to the small reduction in mass due to the absence of microlenses. Hence, if we are interested in the probability of having magnification $\mu_{\rm 3m}>100$, this is basically determined by the millilens and the macromodel. In this situation, microlenses play the role of providing the temporary boost in flux to the  lensed stars moving across the dense web of microcaustics to promote them beyond the detection limit and hence appear as transients. The problem can then be reduced to studying the contribution from a population of millilenses to the probability of having  $\mu_{\rm 2m}>100$ and across an area in the image plane where the macromodel takes different values of $\mu_{\rm 1m}$. 

\section{Scaling with millilens mass and macromodel magnification}
\label{Sect_Scaling}
Having established that the most interesting 3M-lensing effects concentrate around the cusps of the millilenses, and that we can reduce the problem we seek to solve to computing the probability that the macrolens plus millilens produce magnification greater than some value $\mu_{\rm 2m}$, we now focus on the scaling of $A(>\mu_{\rm 2m})$ with the mass of the millilens ($M_{\rm mil}$) and macromodel magnification ($\mu_{\rm 1m}$).  

\begin{figure} 
   \includegraphics[width=8.6cm]{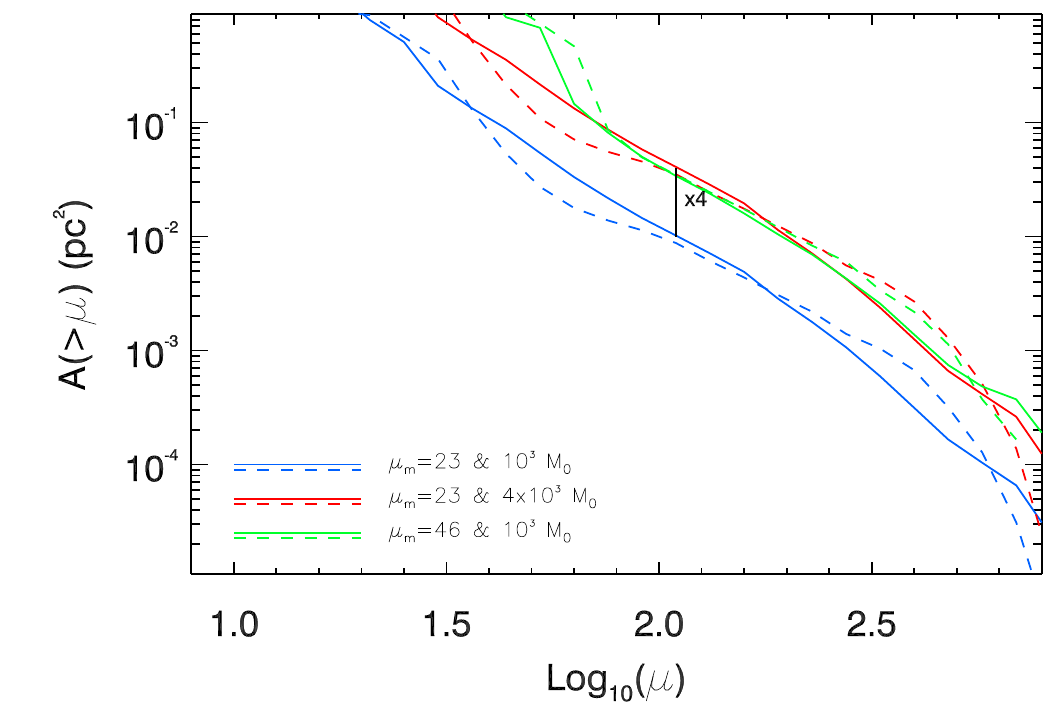}
      \caption{Scaling of probability of magnification. The curves show the area in the source plane with magnification greater than a certain value due to millilenses in the lens plane. Solid lines are for millilenses in regions of the lens plane where the macromodel magnification is positive (positive parity), while dashed lines are for millilenses in regions of the lens plane with negative macromodel magnification (negative parity). Blue curves are for a millilens with mass $10^3\, \Msun$ and macromodel magnification $\pm 23$, red curves are for millilenses with mass  $4\times10^3\, \Msun$ and macromodel magnification $\pm 23$. 
      Green curves are for a millilens with mass $10^3\, \Msun$ and macromodel magnification $\pm 46$. The black vertical line indicates a factor of 4 difference. The probability of magnification scales linearly with the mass of the millilens and quadratically with the macromodel magnification.  
         }
         \label{Fig_Histo2}
\end{figure}

 We characterize this probability by fitting the tail of the magnification with the canonical law $A(>\mu) = A_o/\mu^2$. The parameter $A_o$ defines the strength of the millilens and contains the scaling we seek. 
 Figure~\ref{Fig_Histo2} shows an example with two masses for the millilens and two values for the macromodel magnification. As in Figure~\ref{Fig_Histo1}, dashed lines indicate negative parity and solid lines are for positive parity. 
 The black vertical line shows a multiplicative factor of 4. This factor corresponds to the difference in mass and to the square of the difference in macromodel magnifications. Hence, the parameter $A_o$ scales with mass as $A_o\propto M$ and with macromodel magnification as $A_o\propto \mu_{\rm 1m}^2$. A similar result is found in earlier work for microlenses \citep{Diego2018,Palencia2023}. 

 By fitting the different curves we find the scaling of the probability with the millilens mass ($M_{\rm mil}$) and macromodel magnification ($\mu_{\rm 1m}$),
 \begin{equation}
 A_{\rm 2m}(>\mu) = 0.19 \left( \frac{M_{\rm mil}}{10^3\, \Msun}\right)\left( \frac{\mu_{\rm 1m}}{\mu} \right)^2\, {\rm pc}^2\, .
 \label{Eq_AGTmu}
 \end{equation}
\noindent
 This scaling is almost insensitive to the particular values of the tangential and radial values of the macromodel magnification, and the probability depends only on their product, or $\mu_{\rm 1m}$. In rare  situations where $\mu_r \approx \mu_t$, the caustic shape morphs into a singular point, but the probability of magnification is still given by the same scaling and depends only on the product $\mu_t\times\mu_r=\mu_{\rm 1m}$.
 The law above is derived for the redshifts of the Dragon arc and cluster A370, but it can be rescaled for other redshifts simply by correcting for the factor $D_{ds}/D_dD_s$. 
 Also, the scaling in Equation \ref{Eq_AGTmu} appears to work for individual microlenses. We tested the scaling with a single simulation of a $2\, \Msun$ microlens in a potential with $\mu_{\rm 1m} = 20$ at a resolution of 2\,nas per pixel, and the scaling in Equation \ref{Eq_AGTmu} holds even at this low mass. 
 One can even extrapolate this relation to cluster-scale lenses by considering $\mu_{\rm 1m} = 1$, since cluster lenses are generally in large-scale potentials with magnification $\mu_{\rm 1m} \approx 1$. The prediction for the area above $\mu=30$ for a cluster at $z=0.375$ with mass $10^{15}\, \Msun$, a source at $z=0.725$, and $\mu_{\rm 1m}=1$ is 
 $A(\mu>30) \approx 210$\,kpc$^2$, while for five well-modeled clusters in \cite{Vega-Ferrero2019}  with masses $\sim 10^{15}\, \Msun$ (excluding the supermassive MACS0717 cluster), the area $A(>\mu=30)$ for these redshifts ranges between $\sim 1300$\,kpc$^2$ and $\sim 3700$\,kpc$^2$, corresponding to a factor of $\sim 6$ to $\sim 20$ more. Despite this disagreement, it is still remarkable that the prediction comes to within one order of magnitude, considering there is a 12 orders of magnitude difference in mass between a small $10^3\, \Msun$ millilens and a massive $10^{15}\, \Msun$ galaxy cluster, and the latter are highly irregular, rich in substructure, and with shallower potentials (that are more efficient at increasing the area in the source plane with high magnification).

\section{Probability of 3M-lensing far from the cluster CCs from a population of millilenses.}
\label{Sect_Stat2}
Evolved GCs have masses in the range $\sim10^3$--$10^6\,{\rm M}_\odot$ and are baryon dominated with mass-to-light ratios of a few  \citep{Goudfrooij2016,Harris2017,Bragaglia2017,Baumgardt2018}. Puffy or low-mass GCs are less resilient against disruption from tidal forces in the galaxy cluster, which together with two-body interactions can lead to their complete dissolution.
Almost the entire range of GC luminosities has been measured in the Virgo and Fornax Cluster galaxies \citep{jordan+2007,villegas+2010}, where it is found that the luminosity functions (LFs) of evolved GCs are well matched by a log-normal distribution \cite{Harris2014}.
At higher redshifts, it is expected that the faint end of the LF will be boosted, since young low-mass clusters will not have been disrupted yet \citep{ReinaCampos2022}.
Dynamical disruption mechanisms also affect massive clusters, thus lowering the maximum mass, but the presence of ultracompact dwarf galaxies (UCDs) in the observed samples would prevent detecting differences in this regime. Since colors and luminosities alone are not sufficient to disentangle these two populations, and both would produce the millilensing effect considered in this paper, we consider them both indistinctly.
Recent work based on {\it JWST} has revealed a population of massive GC-like objects in galaxy cluster environments at intermediate redshifts, $z \approx 0.2$--0.4 \citep{Faisst2022,Lee2022,Harris2023}. The high masses of some of these objects, exceeding in some cases $10^7\, \Msun$, are larger than those for massive GCs and are suspected to be the stripped galactic cores of dwarf galaxies \citep{Faisst2022}. The population of GC-like objects in galaxy clusters is then probably a combination of true GCs and UCDs. 

\begin{figure} 
   \includegraphics[width=8.6cm]{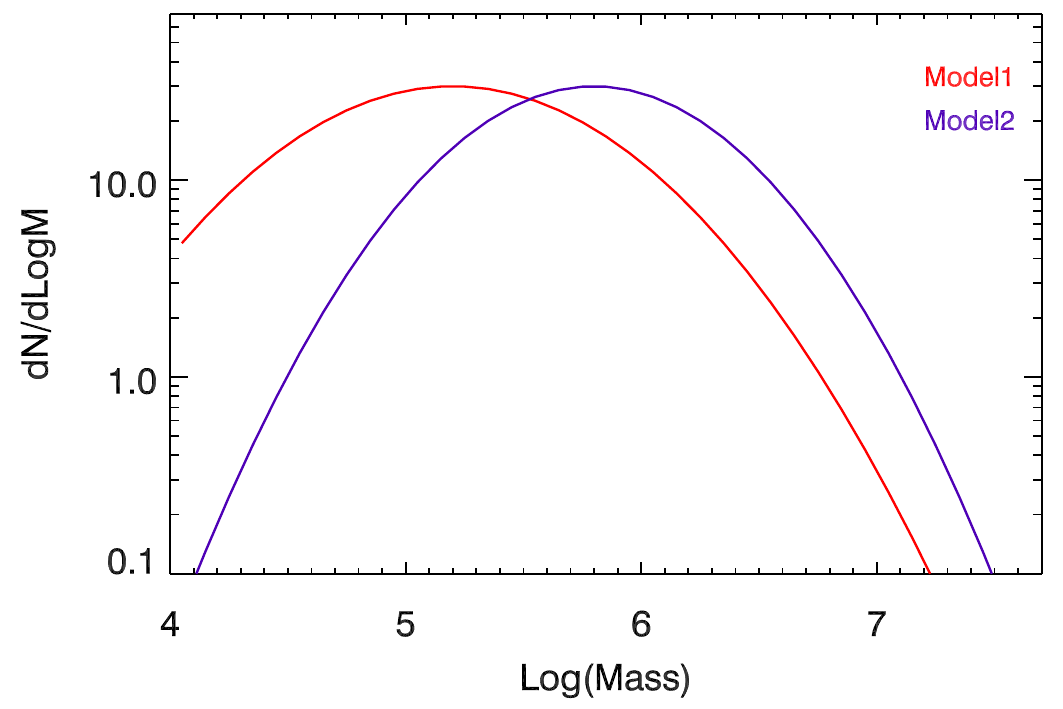}
      \caption{GC mass function. The two colored lines show the two log-normal models used in this work (see text).  
         }
         \label{Fig_GC_massfunction}
\end{figure}

To describe the mass function of GCs, we adopt a log-normal LF \citep{Harris2014, Harris2023}. Assuming a constant mass-to-light ratio, the mass function should be similar to the LF (given as a function of magnitude in that reference). For the log-normal shape, we assume three parameters: (i) the peak, $M_o$, of the log-normal, which depends on the effect of dynamical disruption processes, as well as on the detection of the faintest and harder to detect GCs, (ii) the dispersion, $\sigma$, of the log-normal, and (iii) the number of GCs which we parameterize as a number density of GCs (the total area covered in the lens plane by the Dragon arc is $\sim 1150$\,kpc$^2$, out of which 960\,kpc$^2$ are in the far region). The GC mass function takes the form \citep[see Eq. (1) of ][]{Harris2014}
\begin{equation}
\frac{{\rm d}N}{{\rm d}\log_{10} M} = \frac{N}{\sqrt{2\pi}\sigma}\exp\left[- \frac{(\log_{10}M-\log_{10}M_o)^2}{2\sigma^2} \right],
\end{equation}
where $N$ is a normalization constant. We consider two alternative models that are shown in Figure~\ref{Fig_GC_massfunction}; each one  has a different value of $M_o$ and $\sigma$. Model 1 (with $\log_{10}(M_o)=5.2$ and $\sigma=0.6$) is our reference model and corresponds to the expected mass function of GCs from numerical simulations of star-cluster populations within cosmological zoom-in Milky-Way-mass simulations \citep{ReinaCampos2022}. In contrast, Model 2 (with $\log_{10}(M_o)=5.8$ and $\sigma=0.5$) is an alternative and top-heavy mass function that we use to check the dependency of our results with the GC mass function. 
The value of $\sigma$ in these models is comparable to the universal value derived for the LF by \cite{Harris2014}. 

We can now combine all ingredients and compute the area in the source plane with magnification $\mu_{\rm 2m}>\mu_{\rm crit}$ created by millilenses in the far region. Above $\mu_{\rm crit}$, microcaustics are constantly overlapping in the source plane and the probability of microlensing saturates at its maximum. As discussed earlier, we adopt $\mu_{\rm crit}=100$, which satisfies the supercritical condition $\Sigma_{\rm eff}=\mu_{crit}\Sigma_{*} \gtrsim \Sigma_{\rm crit}$ when $\Sigma_{*}\approx 50\,\Msun\,{\rm pc}^{-2}$. 

The area in the source plane where microlensing is most likely to take place is computed as the integral over the region in the lens plane with macromodel magnification $\mu_{\rm 1m}<\mu_{\rm crit}=100$ and the mass functions of GCs,
\begin{equation}
A_{\rm far}(\mu_{\rm 2m}>100) = \int d\mu_{\rm 1m}\int \frac{dN}{dM}P(\mu_{\rm 1m})A_{\rm 2m}(>\mu)\, dM\, ,
\label{Eq_Atot}
\end{equation}
where $A_{\rm 2m}(>\mu)$ is given by Equation~\ref{Eq_AGTmu}, the magnification is integrated between 1 and 100, and $P(\mu_{\rm 1m})$ is the probability for the macromodel magnification (or area with magnification $\mu$) in the lens plane. This probability goes as $\mu^{-1}$ when taking logarithmic bins in $\mu$.  That is, we take $P(\mu_{\rm 1m})={\rm d}A/{\rm d}\log(\mu)=P_o/\mu_{\rm 1m}$ and determine $P_o$ with the constraint $\int {\rm d}\mu P(\mu_{\rm 1m}) = 1$.

\section{Expected vs. observed number of transients near millilenses in the far region}
\label{Sect_Results1}

With Equation~\ref{Eq_Atot}, we can compute the area in the source plane around millilenses in the far region with magnification $\mu_{\rm 2m}>100$, or $A_{\rm far}(\mu_{\rm 2m}>100)$, but we want to compare this area with the area in the near region satisfying $\mu_{\rm 1m}>100$,  or $A_{\rm near}(\mu_{\rm 1m}>100)$. Most microlensing events are expected to take place in these two areas. Microlensing can in principle take place with similar probability in both areas, provided the number density of stars is the same in both regions.\footnote{In Section~\ref{Sect_MappingDM} we see how this also depends on the LF of the background stars.}

\begin{figure} 
   \includegraphics[width=8.6cm]{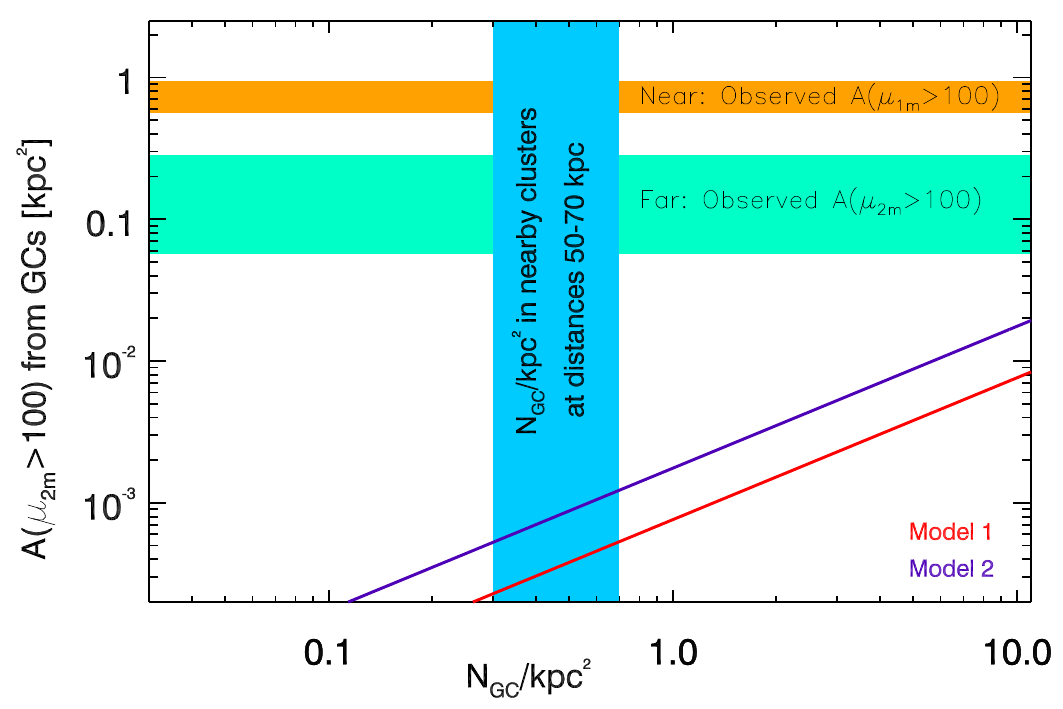}
      \caption{Expected and observed area in the source plane with magnification $\mu>100$. The red and blue solid lines show $A_{\rm far}(\mu_{\rm 2m}>100)$, the expected area in the source plane with magnification greater than 100 around millilenses in the far region for the two different mass functions shown in Figure~\ref{Fig_GC_massfunction}. This area is computed as a function of the number density of millilenses, ${\rm N}_{\rm GC}$, and later rescaled to the area in the far region (960\,kpc$^2$). The blue vertical band shows the typical range of number density of GCs at the distance of the Dragon arc from nearby clusters.  The horizontal orange band shows $A_{\rm near}(\mu_{\rm 1m}>100)$,  the area in the source plane with macromodel magnification $\mu_{\rm 1m}>100$. The green horizontal band represents the fraction of microlensing events found in the far region with respect to the near region ($\sim 0.1$ to $0.5$ times the number of events found in the near region). 
         }
         \label{Fig_Area_vs_NGC}
\end{figure}

The area in the near region is determined by the lens model for the galaxy cluster. We use the free-form WSLAP+ model derived for this cluster with the latest constraints from {\it HST} (see Appendix). 
Based on the WSLAP+ model, we first compute the area in the source plane (from the macromodel) with magnification $> 100$ and that overlaps with the Dragon arc. This area can be computed in the image plane, then divided by a factor of 100 to transform it into source-plane area, and finally divided by an additional factor of 2 to account for the two parities. This results in 0.95\,kpc$^2$ in the near region of the source plane where the macromodel should produce two counterimages with magnification $\mu>100$ each. The two counterimages should appear in the corresponding near region in the image plane (band determined by the two cyan curves in Figure~\ref{Fig_Dragon}). Alternatively, the area above a certain magnification can be computed directly in the source plane with ray-tracing methods. In this case we obtain the total magnification of a source that gets multiply  imaged into $N$ counterimages. At large magnification factors, usually two of the counterimages carry most of the amplification (this happens when the source is very close to a cluster caustic). In this situation one can approximate the total magnification as twice the magnification from each counterimage. To account for this effect we then need to compute the area in the source plane with magnification $> 200$, resulting in an estimate of 0.57\,kpc$^2$ in the source plane.  
Neither method is perfect when addressing global properties of an entire galaxy, especially in the case of the Dragon arc where multiple cluster caustics intersect the background galaxy but the range 0.57--0.95\,kpc$^2$ should be a good approximation to the truth (within a factor of 2).  This range for the area $A_{\rm near}(\mu_{\rm 1m}>100)$ is shown as an orange horizontal band in Figure~\ref{Fig_Area_vs_NGC}. The luminous stars in this area are the most likely to experience microlensing near the cluster CC. 

Before computing the result of Equation~\ref{Eq_Atot}, we confirm that the macromodel probability of the WSLAP+ model does indeed scale as $P(\mu_{\rm 1m})={\rm d}A/{\rm d}\log(\mu)=P_o/\mu_{\rm 1m}$. This is demonstrated in Figure~\ref{Fig_AGTmuDragon} in the Appendix.  
The ordinate in Figure~\ref{Fig_Area_vs_NGC} shows Equation~\ref{Eq_Atot} computed in the far region and for the two GC mass-function models shown in Figure~\ref{Fig_GC_massfunction}. That is, for each model we show the total area near millicaustics in the source plane with magnification greater than $\mu_{\rm crit}=100$, and as a function of the number density of GCs overlapping with the Dragon arc in the far region ($\mu_{\rm 1m}<100$).  Any star in the background galaxy that falls within this area in the source plane will have the same probability of experiencing a microlensing event (creating counterimages in the far region of the image plane) than stars with similar brightness in the near region of the source plane and with an estimated area of 0.57--0.95\,kpc$^2$ (counterimages would form in the near region of the image plane).

The area in the near region (source plane) is shown as a horizontal orange band at the top of the figure. Clearly, the prediction (solid lines) is below the orange band for any reasonable number density of GCs (vertical blue band). 

Microlenses overlapping with the millilenses would increase this only by a small amount since the stellar mass from the ICL overlapping with the millilenses is much smaller (see Figure~\ref{Fig_Histo1}), so the contribution from microlenses overlapping with the millilens to the area, $A(\mu>100)$, is very small.

The abscissa in Figure~\ref{Fig_Area_vs_NGC} shows the number density of GCs. The total number of GCs can be obtained after multiplying by the area contained in the far region of the Dragon arc in the lens plane (960\,kpc$^2$). 
The average mass of a GC after integrating the GC mass function (normalized to $\int dN/dM =1$ GC) is close to the peak of the log-normal,  
so the abscissa also can be transformed into surface mass density by simply multiplying by this number. 
Since the Dragon arc is at a distance of $\sim 50$--70\,kpc  from the brightest cluster galaxy (BCG), we can compare this number density with the one observed in nearby clusters (such as Coma or Virgo). 
The blue vertical band in Figure~\ref{Fig_Area_vs_NGC} is the observed number density in the local universe from \cite{Peng2011} and for distances in the range 50--70\,kpc.

 Recent observations from the {\it HST} Flashlights program  suggest that the observed rate in the far region is almost comparable to the number of events in the near region. The Dragon arc holds the record for the largest number of transient events reported so far in an individual galaxy. \cite{Kelly2022} find seven transients in this arc after comparing two deep epochs in very wide filters taken with {\it HST}. Six of these events have estimated macromodel magnifications below 100 (from two lens models), indicating a clear preference for these events to appear in regions of the lens model where the macromodel magnification is not extreme. The uncertainty in the magnification of these events is relatively high, especially near the CCs, but even adopting a more conservative value for the critical magnification of $\mu_{\rm crit}=30$, three of the events have magnifications below 30 in the two lens models considered by \cite{Kelly2022}. From our lens model, at least two events are clearly in the far region (see Figure~\ref{Fig_Dragon}). As a conservative and generous range, we assume that the ratio of far-to-near events is between 0.1 and 0.5 times the lower bound of the orange band. This range is represented by the green band in  Figure~\ref{Fig_Area_vs_NGC}.

A similar result is found in the Warhol galaxy ($z=0.94$) but with {\it JWST} observations \citep{Yan2023}. Seven transient events were found, with three in regions having macromodel magnification below 100 (and as low as $\mu \approx 30$). Interestingly, all three events peak their emission at wavelengths $\lambda_{\rm peak} > 2$\,$\mu$m, suggesting these are cool stars. For the case of Warhol the rate of far-to-near events would then be close to 0.5, and given the very red nature of these transients, the LBV hypothesis seems less likely. 
The smaller number of events is partially due to the fact that the galaxy is farther away, so it requires even more extreme magnification factors to detect the same star, disfavoring the LBV hypothesis for these events. Also, the cross section of the cluster caustics with the background galaxy is substantially smaller than for the Dragon arc, hence reducing the chance of  finding stars near high-magnification regions. 
On the other hand, Warhol is at half the distance from the BCG than the Dragon arc,  so the number density of GCs and the probability of microlensing near GCs in Warhol should be at least double the probability of the Dragon galaxy, but still far too small to explain the observed ratio of more than 0.1. In the same work \citep{Yan2023}, four additional events are reported in the galaxy Spock, at a slightly larger redshift, $z=1.0054$. One out of the four events was found in a region with predicted macromodel magnification below 100, which would put the rate of far-to-near events at  $\sim 1/3$, again orders of magnitude higher than  expected.
As in the case of Warhol, this transient is also very red, $\lambda_{\rm peak}>2$\,$\mu$m, making the LBV interpretation equally unlikely. 
For the Spock galaxy the high ratio of events far away from the CC is even more striking since this galaxy is in a portion of the cluster with an estimated surface number density of microlenses lower than for Warhol and the Dragon arc, so the amount of magnification needed (from the macromodel) to achieve the critical surface number density is higher. A detailed treatment for the Warhol and Spock galaxies is beyond the scope of this paper \citep[see, however,][where two of the {\it HST} microlensing events in the Spock arc were studied in more detail]{DiegoBUFFALO}. Here we simply use them as additional examples of an apparently high ratio of events in the far to near regions. 


\begin{figure} 
   \includegraphics[width=8.6cm]{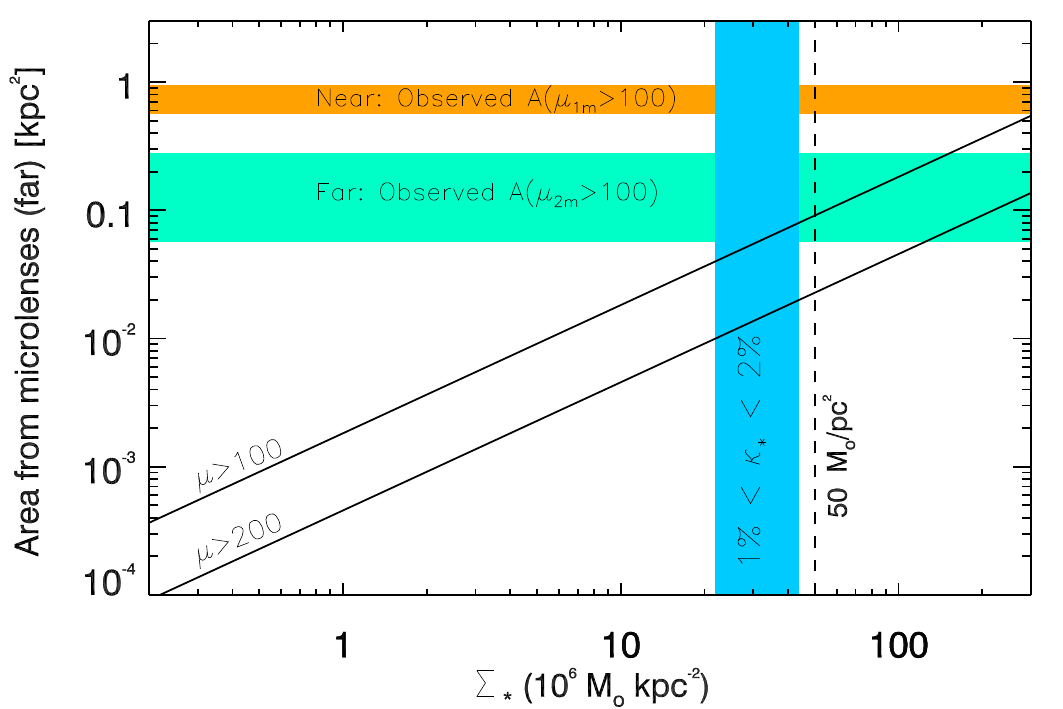}
      \caption{Contribution from microlenses in the far region to high magnification. This result is similar to Figure~\ref{Fig_Area_vs_NGC}, but considers only the effect of macromodel magnification and microlenses in the far region. The two solid black curves show the area in the far region of the source plane where microlenses create magnification factors $> 100$ and 200, and as a function of the surface mass density of microlenses. Above $\mu>100$ in the far region, stars are already very close to a microcaustic and can reach it in a few months. 
      For $\mu>200$ more stars can be detected, but they reach the microcaustic on a shorter timescale. 
      The vertical blue band shows the expected range of surface mass densities for microlenses that constitute 1\% and 2\% of the total projected mass. (The convergence at the redshift of the Dragon arc from the macromodel in the Dragon arc region ranges between 0.58 and 0.62, so we adopt the mean value 0.6, while the shear ranges from 0.36 to 0.38.) The vertical dashed line is the fiducial microlensing model. 
         }
         \label{Fig_Area_vs_micro}
\end{figure}
If the number density of stars that can be detected during a microlensing event is the same in the far and near region, the ratio of the areas in the near and far regions translates directly into the expected rate of microlensing events in the near and far regions. 
In Section~\ref{Sect_MappingDM} we will see how the LF of the background stars plays also an important role in determining the final number of microlensing events, but here we can anticipate that for any reasonable LF, and at GC densities of 1 GC\,kpc$^{-2}$, the area from millilenses in the source plane (and hence the relative probability between the far to near events) is far below what is needed to produce a significant number of microlensing events in the far region (horizontal green band). 
\section{Transients from microlenses alone (no millilenses) in the far region}
\label{Sect_Results2}

So far we have focused all our attention on the possible role played by millilenses at explaining the $0.7 < z < 1$ transient events observed in the far region of cluster CCs. Since in this region the macromodel magnification is relatively small, microcaustics from stars contributing to the ICM do not overlap in the source plane and the probability of microlensing is greatly diminished, but this does not mean microlensing in areas with lower magnification $\mu_{\rm 2m}$ cannot take place. 

One fundamental difference between microlenses and millilenses is that microlenses have a much higher number density. At the distance from the BCG of the Dragon arc, the surface mass density of microlenses in our fiducial model is $\Sigma_{*}=50\, \Msunpc^{-2}$. This estimate is consistent with measurements based on the ICL at similar distances \citep{Montes2022}. In the 960\,kpc$^2$ occupied by the Dragon arc, this surface mass density translates into a total mass of $4.8\times10^{10}\, \Msun$. 
This is a factor of $50$ larger than the mass from GCs assuming a number density of 1 GC per kpc$^2$ and a mean mass of $10^6\, \Msun$ per GC (Model 1).

It is then natural to expect that microlenses alone should play a bigger role than millilenses. We repeat the calculation done for the GCs, but this time as a function of the surface mass density of microlenses and ignoring the contribution from millilenses. Since the GCs assumed earlier are very compact, with masses contained within their effective Einstein radius, they behave as point masses, so we can use the scaling law in Equation~\ref{Eq_AGTmu} by simply replacing the millilens mass by the surface mass density of microlenses. This extrapolation can be tested against the simulation result shown in Figure~\ref{Fig_Histo1}, where for $77\, \Msun$ we find an area above $\mu=100$ of $\sim 5\times10^{-4}$\,pc$^2$, while for the same mass and the scaling in  Equation~\ref{Eq_AGTmu} we expect $7.7\times10^{-4}$\,pc$^2$ (in both cases $\mu_{\rm 1m}=23)$. 

The result for microlenses alone is shown in Figure~\ref{Fig_Area_vs_micro}, where we compare the area in the far region of the source plane with magnifications $\mu>100$ and $\mu>200$. As expected from their larger surface mass density, the contribution from microlenses is substantially more than from millilenses. In the figure, we mark with a vertical dashed line the surface mass density of our fiducial model. The microlenses in this model are sufficient to explain the elevated rate of events in the far region. 
The blue vertical band marks the range of surface mass densities corresponding to convergence from the stellar component between 1\% and 2\% of the total convergence of the cluster at the position of the Dragon arc.  Our fiducial model corresponds to $\kappa_{*}=2.3\%$, a reasonable value for distances between 50 and 70\,kpc to the center of the cluster. 

The results presented so far have not taken into account the LF of the background stars, since we are simply looking at the ratio of areas (or relative probabilities) between the far and near regions where $\mu_{\rm 2m}>100$ (or $\mu_{\rm 3m}>100$ for the far region) and $\mu_{\rm 1m}>100$ (for the near region). The probability of microlensing is proportional to these areas, but the number of stars that can be detected through microlensing (as mentioned earlier, we refer to this group of stars as DTM stars) depends strongly on the LF as we shall see in the next section, where we also discuss the key elements that makes imaging DM substructure with lensed stars possible.

\section{Mapping dark matter substructures with microlensing events}
\label{Sect_MappingDM}
We have seen how millilenses are not the most likely explanation for the high fraction of events (proportional to the area with high magnification) found in the far region, but microlenses (and LBVs) offer a more likely explanation. However, we have also seen how the rate of microlensing events is enhanced around millilenses (or in general perturbations in the small-scale distribution of mass). 
Figures 7 and 15 of \cite{Williams2024} show that the number of highly magnified images is proportional to the length of millilens CCs.
This offers the interesting prospect of using distant stars as backlights and microlensing events as the markers of substructure that is influencing the number of detected microlensing events. We can then map the location of microlensing events, and use them to learn about the distribution of matter along the line of sight. This is analogous to using a photographic plate to trace the distribution of photons crossing an imperfect glass with nonuniform thickness, where in our case the photons are the distant stars being microlensed and the irregularities in the glass screen are the small perturbations in the lensing potential from DM substructures. The analogy with photographic plates will be made more evident later in this section.

So far we have ignored the role played by the LF of the population of lensed stars, but the distribution of microlensing events depends on the distribution of matter and the specific form of the LF. 
Many of the microlensed stars are $\sim 1$--2 (apparent) magnitudes below the detection threshold before microlensing. 
In the following sections, we assume all microlensing events provide a boost of $\sim 2$ mag (on average), so the DTM stars would be the ones that during a microlensing event can be detected. The specific amount of magnification provided by microlenses is irrelevant for our calculations. All that matters is that microlensing can promote fainter stars beyond the detection threshold and make them vary in flux between two epochs, so they can be recognized as transients in the difference of images taken with the same filter. Most of the DTM stars would be undetectable without microlensing, but some may be already detectable with just the boost provided by the macrolens (and the millilens, if one happens to be nearby) and before microlensing, but all of them would appear as transients during a microlensing event. 

\subsection{Number density of DTM stars}
The number density of DTM stars, and in a region with magnification $\mu$, is given by 
\begin{equation}
\rho(\mu,\beta)=\frac{A_s(\mu)}{A_i(\mu)}\int_{L_1(\mu)}^{L_{\rm max}} \phi(L)\, dL \, ,
\label{Eq_rho1}
\end{equation}
where $\phi(L)\propto(1/L)^\beta$ is the classic (per unit area) LF of the background population of stars, $L_1(\mu)=L_{\rm min}/\mu$ with $L_{\rm min}$ the minimum luminosity that could be detected at the redshift of the background galaxy at magnification $\mu=1$ and is set by the depth of the observations, $L_{\rm min}=10^{0.4(m_{\rm thr}-dm(z))}$, with $m_{\rm thr}$ the limiting magnitude of the observations, $dm(z)$ the distance modulus to redshift $z$, and for simplicity we ignore color corrections. In reality, since we are interested in stars that can be detected through microlensing, $L_{\rm min}$ is smaller by a factor of $10^{2/2.5}=6.31$, or 2~mag fainter, and during a microlensing event, a star with luminosity $L_{\rm min}/6.31$ will be magnified enough to be detected.  
Since these are DTM stars, they can be detected when experiencing a microlensing event, and eventually all of them would be detected if one could monitor the area for a sufficiently long  time.  
$L_{\rm max}$ is the most luminous star in the area considered and depends on the assumed shape of the LF, or the existence of a limiting luminosity for the stars, such as the Humphreys-Davidson (HD) limit for RSGs s\citep{Humphreys1978}. For our purposes, we assume the observations are deep enough such that we can see microlensing from stars much less luminous (before magnification) than the most luminous star in the portion of the galaxy being magnified. In the expression above we assume that the probability of magnification when millilenses and microlenses are added is similar to the one given by the macromodel alone, so we can simply rely on the macromodel magnification. This is a very good approximation since millilenses and microlenses do not modify the probability of magnification significantly (when computed over areas much larger than the scale of the micro or millilenses), but rather they borrow magnification from surrounding regions and redistribute it around the millilenses and microlenses \citep[see Figures7 and 8 of ][where the probabilities of magnification for the smooth model and the smooth model plus microlenses are very similar]{DiegoExtreme}. 
For moderate magnification factors and typical depths, $L_1(\mu)>L_{\rm max}$ and the integral is zero in Equation~\ref{Eq_rho1}, but for sufficiently large $\mu$, $L_1(\mu)<L_{\rm max}$ and the number of DTM stars (and consequently microlensing detections) is greater than zero. The areas in the image and source plane are $A_i$ and $A_s$ (respectively), and they are related by $A_s=A_i/\mu$. Taking as unit area $A_s=1$ and replacing $\phi(L)$ by $(1/L)^{\beta}$, we find

\begin{equation}
\rho(\mu,\beta)= \frac{L^{1-\beta}}{\mu(1-\beta)}\bigg|^{L_{\rm max}}_{L_1(\mu)}
\end{equation}
except for if
$\beta=1$, 
in which case the number density is
\begin{equation}
\rho(\mu,\beta)= \frac{\log(L)}{\mu}\bigg|^{L_{\rm max}}_{L_1(\mu)}\, .
\end{equation}
For a steep LF with $\beta=3$ and sufficiently large values of $\mu$, we find that $\rho(\mu) \propto \mu$, where we have assumed that $L_1(\mu) << L_{\rm max}$. The number density of DTM stars from such  a population of background stars would then directly trace the magnification, and with it the distribution of mass on small scales. If microlenses are present in this area, they will make these DTM stars detectable, with a rate of microlensing events that increases with the abundance of microlenses as shown in Section~\ref{Sect_Results2}. The tight relation between the LF and the number of microlensing events is discussed later in Section \ref{Subsect_LFandMicro}. 

In general, for any $\beta>1$ and for $L_{\rm min}/\mu << L_{\rm max}$, the number density of DTM stars  scales with $\mu$ as
\begin{equation}
\rho(\mu,\beta) \propto \mu^{(\beta-2)} \, .
\label{Eq_rho2}
\end{equation}
\noindent
Based on this, we can define two distinct regimes, which we identify with traditional photographic plate imaging, where photons are crossing a glass with nonuniform thickness. For steep LF with $\beta>2$, we are in the positive-imaging regime. Here, the number density of DTM stars (the photons that reach the photographic plate) is larger around substructures with larger magnification factors (or in our analogy, when photons are crossing portions of the glass that focus the light more into the photographic plate). For shallow LF with $\beta<2$ we are in the negative-imaging regime, where the number density of DTM stars is reduced around substructures. In our photographic-plate analogy, this would correspond to the negative of the photograhic plate, where the silver particles have absorbed more photons behind the small-scale structures. For the particular case of an LF with $\beta = 2$, we expect the number density of events to be uniform and independent of magnification. The photographic analogy would be the superposition of the positive image and the negative plate, leaving as a result a homogeneous image.

To better illustrate this specific case, we discuss a simple experiment.  
For $\beta=2$, and considering two logarithmic bins in magnification, we can think of two regions, $A$ and $B$, with the same LF but different magnifications; 
$10<\mu_B<100$ and $100<\mu_A<1000$. 
Region $A$ has mean magnification a factor of 10 larger than region $B$. In the image plane, region $A$ is 10 times smaller than region $B$ (true for logarithmic bins in $\mu$), while in the source plane, region $A$ is $10^2$ times smaller than region $B$. The number density of DTM stars in the source plane of region $A$ is 10 times larger than in region B ($\int_{L_1}^{\infty} L^{-2}\, dL \propto L_{1}^{-1}$, with $L_1$ 10 times smaller in $A$ than in $B$), but the area is $10^2$ smaller so there are 10 times fewer objects detected in  the image plane in region $A$ than in $B$ (see Figure~\ref{Fig_LumFunct}), but since the area in the image plane of $B$ is 10 times larger, the number density of DTM stars in the image plane is the same in $A$ and $B$. 

The case of a shallow LF, $\beta<2$, is counterintuitive since we expect to see a smaller number density of DTM stars  in regions of higher magnification. This shallow LF resembles the faint end of the LF of quasistellar objects (QSOs) at high redshift. Most lensed QSOs are found in regions with moderate magnification factors, in agreement with Equation~\ref{Eq_rho2} above. This is similar to the enhancement-dilution effect, or magnification bias, discussed in the context of distant lensed galaxies and QSOs \citep{Canizares1981,Narayan1989,Borgeest1991,Narayan1993,Broadhurst1995,Umetsu2014}.

\subsection{The observed luminosity function.}
The LF of the observed events, $\hat{\phi}(L)$, can be directly related to the LF of the background stars (before magnification). For an LF with $\beta\neq3$ and considering a region with minimum magnification $\mu_1$, we have 
\begin{equation}
\hat{\phi}(L)= \int_{\mu_1}^{\mu_{\rm max}} \phi(L/\mu)A_s(\mu)\, \frac{d\mu}{\mu} \propto \mu_1^{\beta-3}\phi(L)\, ,
\label{Eq_LensedPhi}
\end{equation}
where $\mu_{\rm max}$ is the maximum magnification for a star, $\mu_{\rm max}\approx 10^4$ for supergiant stars (see Section~\ref{subsect_duration}), and we have assumed $\mu_{\rm max}>>\mu_1$.  The extra term $1/\mu$ inside the integral is the reduction in luminosity bin size at magnification $\mu$, and  $A_s(\mu)\propto\mu^{-3}$. 
For the particular case of $\beta=3$ we have

\begin{equation}
\hat{\phi}(L) \propto \left({\rm log(\mu_{\rm max})-{\rm log}(\mu_1)} \right)\times\phi(L)\, ,
\label{Eq_LensedPhi2}
\end{equation}
which has a weak dependence on $\mu_1$. 
As discussed in Section~\ref{Sect_Def}, the luminosity functions in Equations~\ref{Eq_LensedPhi} and~\ref{Eq_LensedPhi2} do not conform with the classic definition of number density per luminosity and unit area, but instead correspond to regions in the image plane with an area that depends on $\mu_1$ and $\mu_{\rm max}$. 
The LF maintains its form, but its amplitude (compared with the amplitude of $\phi(L$)), scales as $\mu_1^{\beta-3}$ for $\beta\neq 3$, and remains virtually independent of the magnification when $\beta\approx 3$. For a LF with $\beta\approx3$, the reduction in area in the source plane ($A_s(\mu) \propto \mu^{-3}$) is almost perfectly compensated by the increase in the number of objects with smaller luminosity $L/\mu$. In this case, the number of lensed objects per logarithmic interval in magnification is the same at all magnifications (a visual example of this constancy in $\mu$ is shown in Section~\ref{subsect_MonteCarlo}). 

Similarly, we can define the probability of magnification of the lensed stars, $\hat{\phi({\mu})}$ as
\begin{equation}
\hat{\phi}(\mu)= \int_{L_{\rm min}}^{\infty} \phi(L/\mu)A_s(\mu)\, dL  \propto \frac{\mu^{\beta-3}}{L_{\rm min}^{\beta-1}}\, .
\label{Eq_LensedPhiMu}
\end{equation}
For an LF with $\beta = 3$, all magnifications have similar probability, so the observed population of lensed stars will have equal fractions of fainter and luminous stars. A shallower LF with $\beta<3$, will be dominated by low-magnification events far from the CC (i.e., from intrinsically very luminous stars), 
while a steeper LF with $\beta>3$ will produce mostly 
high-magnification events near the CC (i.e., low intrinsic luminosity stars).

\subsection{Connecting the distribution of transients with the LF and the amount of substructure.}
\label{Subsect_LFandMicro}
Interestingly, the observed number density of  events in Flashlights traces the magnification (higher concentration of events in the near region), so this points (in principle) to a population of DTM with $\beta>2$ (Equation~\ref{Eq_rho2}). 
But this would be true only if substructure is not present. The relation between the LF and substructure adds complexity to this interpretation, as we have seen in Sections~\ref{Sect_Results1} and~\ref{Sect_Results2}, where the probability of having transient events (proportional to the area where large magnification factors are possible) in the far region depends also on the amount of substructure (micro- and millilenses). To estimate the ratio of microlensing events in the far and near regions, one needs to take into account both the LF and the amount of substructure. 


We begin by computing the number of DTM stars in a logarithmic bin in magnification. For the sake of clarity, we derive the scaling with $\mu$ both in the image and source planes, and show how they are both equivalent. To show this scaling with $\mu$, we consider two areas, $A$ and $B$, with magnifications $\mu_A<\mu_B$. For simplicity, the widths in magnification of areas $A$ and $B$ are the same in logarithmic scale. In particular, we will be considering two bins in magnification (in log scale), $10<\mu<100$ for the far region and $100<\mu<1000$ for the near region. 
When considering logarithmic bins, the area of $A$ is $\mu_B/\mu_A$ times bigger than the area of $B$ when computed in the image plane. In the source plane, the area of $B$ is reduced in size by an extra factor $\mu_B/\mu_A$ since $A_s=A_i/\mu$. That is, when computing areas in the source plane the area of $A$ is $(\mu_B/\mu_A)^2$ times bigger than the area of $B$ as expected \citep{Schneider1992}.

\noindent
{\bf Image plane interpretation:}
We count DTM stars in the image plane in areas $A$ and $B$ with magnification $\mu_A$ and $\mu_B$. The number of DTM stars are the ones that are found in the smaller areas $A/\mu_A$ and $B/\mu_B$ in the source plane, and above the luminosity  $L_1(\mu)=L_{\rm min}/\mu$, where $\mu=\mu_A$ or $\mu=\mu_B$. Since $B=A\mu_A/\mu_B$, then the area of $B$ in the source plane is $B/\mu_B\propto A/\mu_B^2$, and the number of DTM stars scales with $\mu$ as 
\begin{equation}
\frac{dN_{DTM}}{d{\rm log}(\mu)} \propto \frac{1}{\mu^2} \int_{L_1(\mu)}^{L_{\rm max}}\phi(L)dL  
\propto \frac{\mu^{\beta-3}}{L_{\rm min}^{\beta-1}}\, .
\label{Eq_NDTM1}
\end{equation}

\noindent
{\bf Source plane interpretation:}
We count stars that fall in the source plane in areas $A$ and $B$ with magnifications $\mu_A$ and $\mu_B$. In this case, the calculation is simplified since we can work directly with the area in the source plane which scales as $\propto \mu^{-2}$. 
The number of DTM stars above luminosities $L_1(\mu)=L_{min}/\mu$ is then
\begin{equation}
\frac{dN_{DTM}}{d{\rm log}(\mu)} \propto \frac{1}{\mu^2} \int_{L_1(\mu)}^{L_{\rm max}}\phi(L)dL,   
\propto \frac{\mu^{\beta-3}}{L_{\rm min}^{\beta-1}}\, ,
\label{Eq_NDTM}
\end{equation}
and thus equivalent to Equation~\ref{Eq_NDTM1}. In the above equations, where we have made the usual approach that $L_{\rm max} >> L_1(\mu) = L_{\rm min}/\mu$. For DTM stars, we have seen how $L_{\rm min}$ is approximately 6.31 times below the luminosity corresponding to the detection limit ($2.5{\rm log}_{10}(6.31)\approx 2$ mag). We have also ignored the multiplicity of counterimages, but this cancels out when considering the ratio of events in the near and far region, assuming the multiplicity is the same in both regions.     
From Equation~\ref{Eq_NDTM}, and for an LF with $\beta=3$, we expect the same number of DTM stars per logarithmic bin in magnification (see also Equation~\ref{Eq_LensedPhiMu}). In the image plane, the area per logarithmic bin in magnification scales as $\mu^{-1}$, so the number density of DTM stars for this case would go as $\mu$ and trace the magnification, in agreement with Equation~\ref{Eq_rho2}.

The total number of stars that experience microlensing in the far region is given by the number of DTM stars in that region (Eq.~\ref{Eq_NDTM}) times the probability of each star to experience microlensing. This probability is proportional to the black solid line in Figure~\ref{Fig_Area_vs_micro}:
\begin{equation}
N_{\rm far} \propto  \frac{dN_{\rm DTM}}{d{\rm log}(\mu)} \times A(\mu_{\rm 2m}>100,\Sigma_{*})\, .
\end{equation}
For the near region, we have a similar expression, but replacing $A(\mu_{\rm 2m}>100,\Sigma_{*})$ by $A(\mu_{\rm 1m}>100)$, which is given by the orange band in Figure~\ref{Fig_Area_vs_micro}. 
Here we ignore millilenses since we have established in Section~\ref{Sect_Results2} that the dominant effect is coming from microlenses.

We can now express the ratio of events in the near and far regions:
\begin{equation}
\frac{N_{\rm near}}{N_{\rm far}}= 
  \left( \frac{\hat{\mu}_{\rm near}}{\hat{\mu}_{\rm far}} \right)^{\beta-3} \frac{A(\mu_{\rm 1m}>100)}{A(\mu_{\rm 2m}>100,\Sigma_{*})}\, .
\end{equation}
In the expression above, we made the simplification that in the far and near regions the number of events can be expressed in terms of their corresponding  average magnification, $\hat{\mu}$, computed as the mean magnification in the source-plane region with $\mu>\mu_1$,
\begin{equation}
\hat{\mu} = \frac{\int_{\mu_1}^{\infty}\mu A_s d\mu}{\int_{\mu_1}^{\infty} A_s d\mu} = 2\mu_1 \, .
\label{Eq_hatmu}
\end{equation}
Here we adopt $\mu_1=10$ for the far region of the Dragon arc and $\mu_1=\mu_{\rm crit}=100$ for the near region.


\begin{figure} 
   \includegraphics[width=9.0cm]{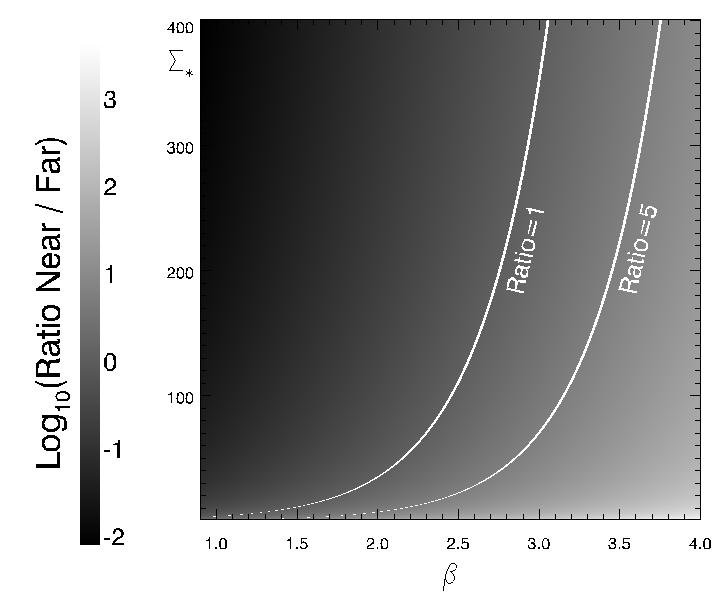}
      \caption{Ratio $N_{\rm near}/N_{\rm far}$ as a function of $\beta$ and $\Sigma_{*}$ (expressed in units of $\Msunpc^{-2)}$. The white lines show the combination of $\beta$ and $\Sigma_{*}$ that predict the same number of events in the far and near regions (Ratio = 1) or five time more events in the near region than in the far region (Ratio = 5). 
        }
         \label{Fig_BetaSigma_2D}
\end{figure}

The value of $\beta$ can be obtained by inverting the equation above, 
\begin{equation}
\beta=3+\frac{{\rm log}(N_{\rm near}/N_{\rm far})-{\rm log}(R_A(\Sigma_{*}))}
{{\rm log}(\hat{\mu}_{\rm near}/\hat{\mu}_{\rm far})}\, ,
\label{Eq_Beta}
\end{equation}
where $R_A$ is the ratio of areas,
\begin{equation}
R_A(\Sigma_{*}) = \frac{A(\mu_{\rm 1m}>100)}{A(\mu_{\rm 2m}>100,\Sigma_{*})}=A(\mu_{\rm 1m}>100)\frac{500}{\Sigma_{*}(\Msunpc^{-2})}\, ,
\end{equation}
where from Figure~\ref{Fig_Area_vs_micro}) we have $A(\mu_{\rm 2m}>100,\Sigma_{*})=\Sigma_{*}(\Msunpc^{-2})/500$. For the near region, we take  $A(\mu_{\rm 1m}>100)\approx 0.7$ kpc$^2$, which is approximately in the middle of the orange band in Figure~\ref{Fig_Area_vs_micro}. That is, $R_A=7$ for $\Sigma_{*} = 50\, \Msunpc^{-2}$. Only the ratio of areas in the far and near regions is relevant for this calculation, and this ratio of areas is independent of the value of $\mu_{\rm crit}$ (see Section~\ref{subsec_mucrit} below). 
The number of microlensing events is  $N_{\rm far}=2$, (numbers 3 and 6 in Figure~\ref{Fig_Dragon}) and $N_{\rm near}=5$, while the ratio $\hat{\mu}_{\rm near}/\hat{\mu}_{\rm far}=10$. Replacing these numbers in Equation~\ref{Eq_Beta}, we finally obtain $\beta=2.55^{+0.69}_{-0.48}\,^{+0.18}_{-0.29} = 2.55^{+0.72}_{-0.56}$, where the first error comes from Poissonian uncertainty in $N_{\rm near}$ and $N_{\rm far}$, and the second error corresponds to the range $25\, \Msunpc^{-2} < \Sigma_{*} < 75\, \Msunpc^{-2}$. The final error bar is obtained after adding in quadrature the first two uncertainties.

With Equation~\ref{Eq_Beta}, $\beta$ can be quickly calculated for any arbitrary amount of substructure. Even though we have expressed  Equation~\ref{Eq_Beta} as a function of the stellar surface mass density, $\Sigma_{*}$, in truth this value represents all substructure that contributes to the area in the source plane with magnification $\mu>100$.  For a larger value of $\Sigma_{*}=140\, \Msunpc^{-2}$ we find $\beta=3$ and for $\Sigma_{*}=14\, \Msunpc^{-2}$ we find $\beta=2$, or $\beta=1$ for an unrealistically low $\Sigma_{*}=1.4\, \Msunpc^{-2}$. Since $\beta$ can be measured directly through the observed LF (when sufficient events are available), one can invert Equation~\ref{Eq_Beta} and derive $\Sigma_{*}$, or in general the surface mass density of substructure since all small substructure contributes to $\Sigma_{*}$.

\begin{figure} 
   \includegraphics[width=9.0cm]{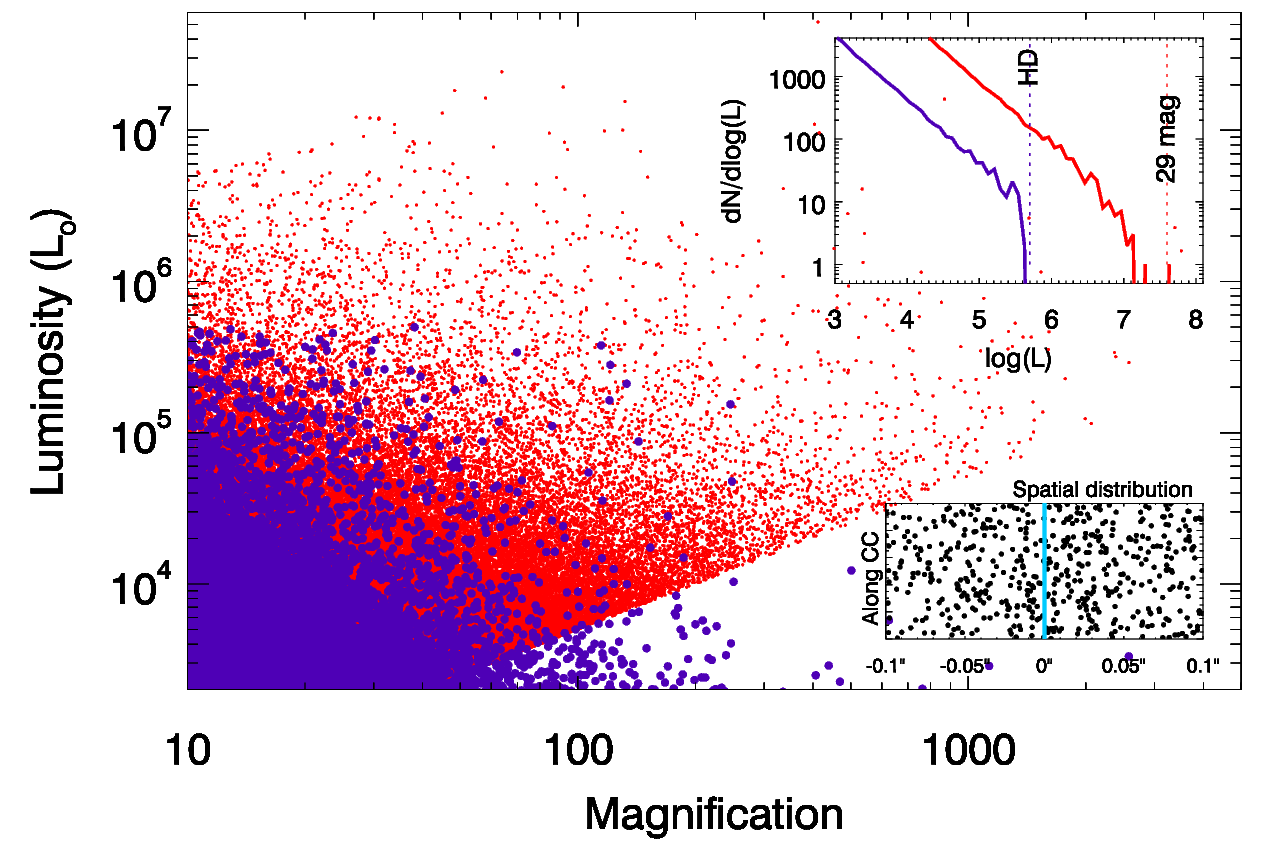}   
      \caption{Monte Carlo realization of lensed stars for a case with $\beta=2$. 
      For this example, half a million stars with luminosities in the range $50\, \Lsun < L < 5\times10^5\, \Lsun$ are placed in a region with magnification $\mu_{\rm 3m}>10$. The maximum luminosity corresponds to the observed HD limit for supergiant stars in the local universe \cite{Humphreys1978}. The diagonal cut in the red points is just $L_{\rm obs}=50\times\mu\, \Lsun$. The simulation is complete above $L_{\rm obs} \approx 10^6\, \Lsun$. 
      The LF of stars  has a slope of $\beta=2$. Blue dots represent stars before magnification and red dots are the magnified stars. In the top-right inset we show the LF of the stars before magnification (blue curve) and the observed LF after magnification (red curve).  In the bottom-right inset we show the spatial distribution of the 500 brightest events and for a model with magnification $\mu=1"/d$ where $d$ is the distance to the CC (marked with a vertical light blue line at $d=0"$) in arcseconds. 
         }
         \label{Fig_LumFunct}
\end{figure}

Equation~\ref{Eq_Beta} summarizes the intricate relationship between the number of observed microlensing events, the LF, and the amount of substructure. The same ratio of events in the far and near regions can be obtained by (i) reducing $\beta$ (hence increasing the relative number of DTM stars in the far region) and reducing $\Sigma_{*}$, or (ii) increasing $\beta$, which increases the number of DTM stars in the near region in relation to the number of available DTM stars in the far region, but increasing $\Sigma_{*}$ as well, thus compensating the reduction of DTM stars in the far region by increasing the chance of a microlensing event.

A visual version of Equation~\ref{Eq_Beta} is shown in Figure~\ref{Fig_BetaSigma_2D}, where we invert the equation to show the ratio of events in the near and far regions as a function of $\beta$ and the amount of substructure $\Sigma_{*}$. 
In the figure we highlight with white lines two possible combinations of the parameters $\Sigma_{*}$ and $\beta$ that produce equal ratios of events in the far and near regions, or five times more events in the near region than in the far region. The measured ratio using Flashlights data (2.5) falls in between these two lines. 
Future observations of this fascinating galaxy will improve the constraints of the observed ratio of events near-to-far and  the exponent  $\beta$. This can later be used to derive the amount of substructure,  $\Sigma_{*}$, needed to make the observed ratio compatible with the observed $\beta$.

\subsection{Validation with Monte Carlo simulations}
\label{subsect_MonteCarlo}

\begin{figure} 
   \includegraphics[width=9.0cm]{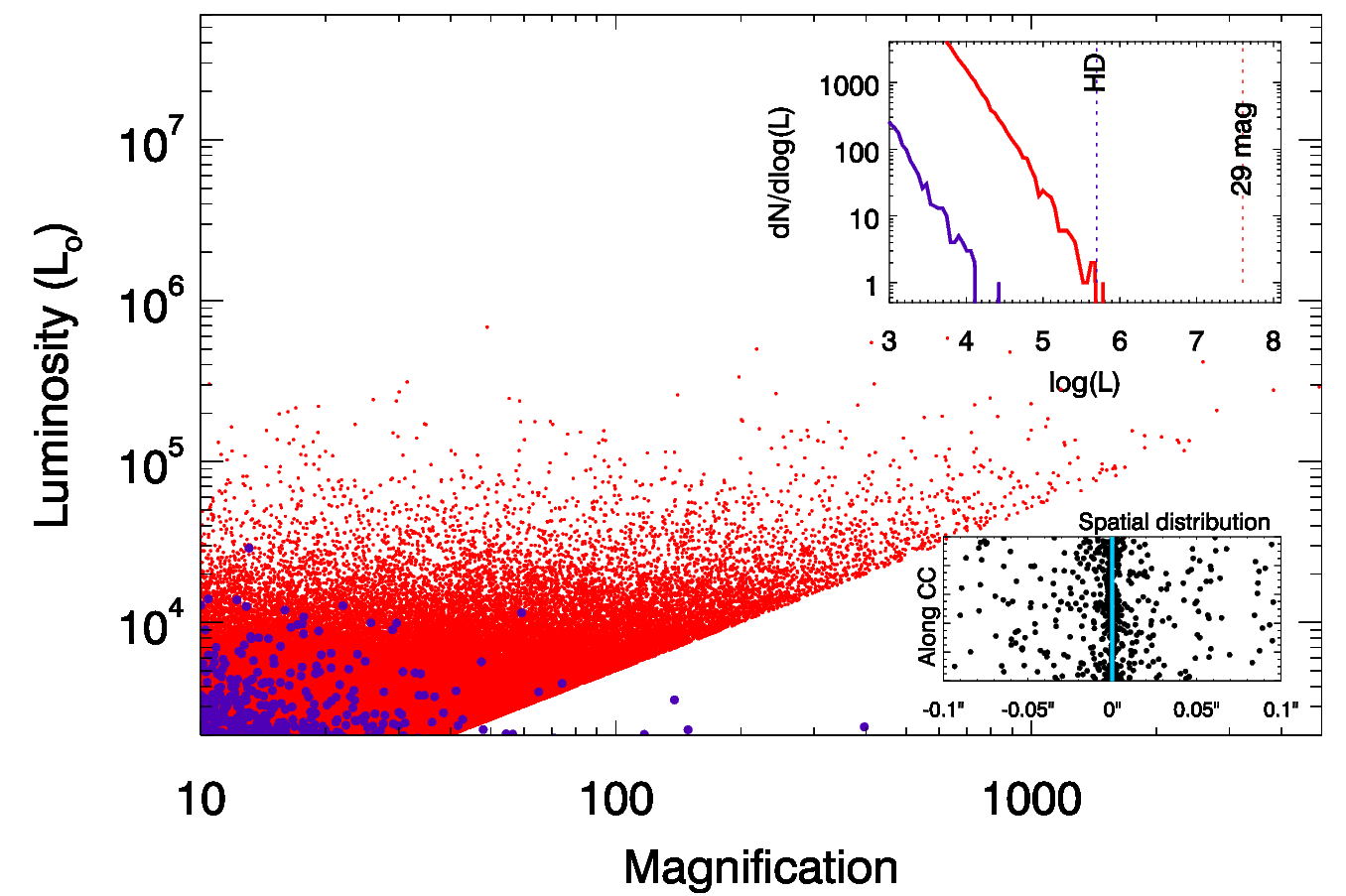}   
      \caption{Similar to Figure~\ref{Fig_LumFunct} but for the case where $\beta=3$. In this case, all magnifications have comparable probability as predicted by Equation~\ref{Eq_LensedPhiMu}. The lensed LF (thick red line) lies a factor of $\hat{\mu}^{\beta-1}$ above the nonlensed LF (blue line). The smaller number of events at high luminosity when compared to Figure~\ref{Fig_LumFunct} is due to the fact that the number of stars in the simulation is the same in both cases, but for the steeper LF there are more stars with lower luminosities. The number density if events concentrates around the CC, tracing the magnification. 
         }
         \label{Fig_LumFunct2}
\end{figure}

In order to test the validity of Equations~\ref{Eq_rho2},~\ref{Eq_LensedPhi},~\ref{Eq_LensedPhiMu}, and \ref{Eq_NDTM}, we perform Monte Carlo simulations \citep[see][ where they also used simulated data to study the particular case of Icarus]{Kelly2018}. 
One example is shown in Figure~\ref{Fig_LumFunct} for the particular case of $\beta=2$ and computed in a region where the magnification is $> 10$ (i.e., $\mu_1 = 10$). We create a sample of half a million stars from the LF with luminosities $L>50\, \Lsun$. This sample is shown as thick blue dots. For this simulation we have set an upper limit to the intrinsic luminosity equal to the HD limit of $5\times10^5\, \Lsun$, so no blue dots are found above this value. The LF from this sample is shown as a blue line in the top-right panel. 
To each star we assign a random magnification in the interval $\mu_1<\mu\mu_{max}$ and following the canonical probability $A_s(\mu)\propto\mu^{-3}$. 
After multiplying the luminosity by the magnification, 
the magnified stars are shown as red dots, with the magnification for each star indicated on the abscissa. We compute the lensed LF from the same sample of half a million stars (red curve in the top-right panel). In a realistic situation, the number of stars in an area with magnification $\mu_1< \mu < \mu_{\rm max}$ should be a factor $\hat{\mu}^2$ smaller,  but we use the same sample for convenience. In these conditions, the amplitude of the red line  scales as $\hat{\mu}^{\beta-1}$, instead of the expected  $\hat{\mu}^{\beta-1}$ from Equation~\ref{Eq_LensedPhi}, so for this particular case of $\beta=1$, the red line is above the blue one by a factor $\hat{mu}=20$. 
Only one star is above a detection threshold of 29\,mag, but a few dozen have apparent luminosities (after magnification) above several million and are in regions where the magnification already exceeds the critical value, $\mu_{\rm crit}=100$, so they are good candidates to move toward a microcaustic and be promoted beyond the detection threshold.

The diagonal cut in the red points corresponds to the smallest luminosity considered in the simulation, $50\, \Lsun$. 
In the simulation, there is no star brighter than the HD limit of $L_{\rm max}\approx 5\times10^5\, \Lsun$, while in the lensed sample, we can reach apparent luminosities exceeding $10^7\,\Lsun$. The most luminous stars in the lensed sample correspond in this case to relatively moderate magnification factors, $\mu \approx 400$ (before microlensing).

Regarding the number density, Equation~\ref{Eq_rho2}, the uniform number density of observed events when $\beta=2$, is also well reproduced by the Monte Carlo, as shown in the bottom-right inset plot of Figure~\ref{Fig_LumFunct}. The abscissa shows the inverse of the magnification of the observed events, which can be transformed into a distance to the CC for an spherical isothermal lens model with a small Einstein radius of $\sim 1''$. A small reduction in the number density is observed at larger distances (small magnifications). This reduction in number is due to the imposed HD limit in the Monte Carlo simulation.  

The dependence on the exponent $\beta$ is made more evident when we compare the previous result with the Monte Carlo simulation for the case with $\beta=3$, and shown in Figure~\ref{Fig_LumFunct2}. The red points have a uniform distribution in magnification, as predicted by Equation~\ref{Eq_LensedPhiMu}. As before, the observed LF (thick red line in the upper-right inset plot) lies above the blue curve by a factor $\hat{\mu}^{\beta-1} = (2\times10)^2$. The spatial distribution shows a much different distribution than in the case with $\beta=2$, with the number density directly tracing the magnification (Equation~\ref{Eq_rho2}).

\subsection{A possible invisible millilens in the Dragon arc}

Since $\beta$ can be estimated directly from the observed $\hat{\phi(L)}$, or from the distribution of events as discussed above, combining the spatial distribution of the number density with the observed LF it is then possible to identify deviations that can be attributed to local departures from the macromodel magnification or regions in the source plane with a different LF. 
Departures from a smooth distribution in the number density can be taken as evidence for substructure, which can locally increase the number density of DTM stars (and transient events) according to  $\rho(\mu,\beta) \propto \mu^{(\beta-2)}$. The number density of microlensing events is then a direct tracer of substructure, and can be used to map the underlying structure of DM fluctuations on subarcsecond scales and down to the milliarcsecond scale \citep[see also the CC structure around millilenses in][Fig.3]{Williams2024}.

In Section~\ref{Sect_Stat1} we discussed how a  millilens with mass as small as $2\times10^3\, \Msun$ can boost the probability of microlensing by at least an order of magnitude when compared to the case of microlenses only (see Figure~\ref{Fig_Histo1}). Such a millilens and its associated Einstein ring would be too small to be resolved even with {\it JWST}, so all the microlensing events near the cusps of the millilens would seem to originate from the same pixel. Since the timescale for a single star with $\mu\approx 100$ to reach the  closest microcaustic is about 1\,yr (see next section), repeated observations with high cadence (weeks) should reveal the population of bright stars behind the microlens as each one crosses one of the multiple microcaustics around the millilens cusp. As discussed above, the LF of all events coming from this single pixel should be proportional to the LF of the background population of luminous stars, and with the same exponent $\beta$.

For larger millilens masses, and $\beta>2$, the microlensing events around the millilens will take place in neighboring pixels and form a cluster of events. The clustering of detected events can be used to trace the underlying mass distribution of millilenses. Since microlensing events are most likely in supercritical regions, if a substructure in the far region becomes supercritical, microlensing events will more likely be detected around that region than in nearby subcritical regions. If enough events are detected in a lensed galaxy, a pattern emerges with clusters of events at the position of these substructures. We can approximate the size of a supercritical region of a substructure with mass $M_{\rm sub}$ as the area contained within its observed Einstein radius, which is given by 
$\Theta_{\rm obs} \approx \sqrt{\mu_{\rm 1m}}\times \Theta_E$  \citep{Diego2018}. 
If the substructure has circular symmetry, $\Theta_E=\sqrt{(4GM_{\rm sub}/c^2)(D_{ds}/(D_dD_s))}$, and the mass of the substructure can be obtained as
\begin{equation}
M_{\rm sub} = \Theta_{\rm obs}^2\frac{c^2}{4G\mu_{\rm 1m}}\frac{D_dD_s}{D_{ds}}\, .
\end{equation}
Events 3 and 6 in Figure~\ref{Fig_Dragon} are located in the far region but separated by only $0.15''$. At this position $\mu_{\rm 1m}=13$ from our lens model. If we assume the observed Einstein radius is half the separation between the events, from the relation above we would obtain a mass for the possible undetected substructure along the line of sight to these events of $\sim 1.3 \times 10^8\, \Msun$ within the Einstein radius of the substructure. The virial mass of the substructure could be significantly higher if it is not concentrated enough to contain most of its mass within its Einstein radius. From the $N$-body simulations discussed in Section~\ref{sect_Satellites},  we expect of order 1 satellite galaxy with this mass and overlapping with the Dragon arc. This is a tantalizing result, but it cannot be taken too seriously because it lacks  statistical significance, and the two events mentioned above could simply be a chance occurrence of two microlensing events that happen to take place near each other. However, if substructures in this mass range exist in the cluster in the far region of the Dragon arc, they will become more evident with future observations of this arc, since new events will have a higher tendency to appear in positions near previous events. Future {\it JWST} observations may also reveal the hidden substructure that is increasing the microlensing rate at this location, or alternatively there is an overdensity of luminous stars at this position in the Dragon arc that can also increase the rate of mcirolensing events there.
LBV can be distinguished from genuine microlensing events through their light curves, since they have light curves that depart from the $1/\sqrt{t-t_o}$ behaviour expected for microlensing events ($t$ is time and $t_o$ is the crossing time if the microcaustic). 


\section{Discussion}
\label{Sect_Discussion}
It is important to discuss some of the approximations in this work and consider aspects that have not been treated in the previous sections but affect some of our conclusions. 

\subsection{Duration of microlensing events}
\label{subsect_duration}

\begin{figure} 
   \includegraphics[width=9.0cm]{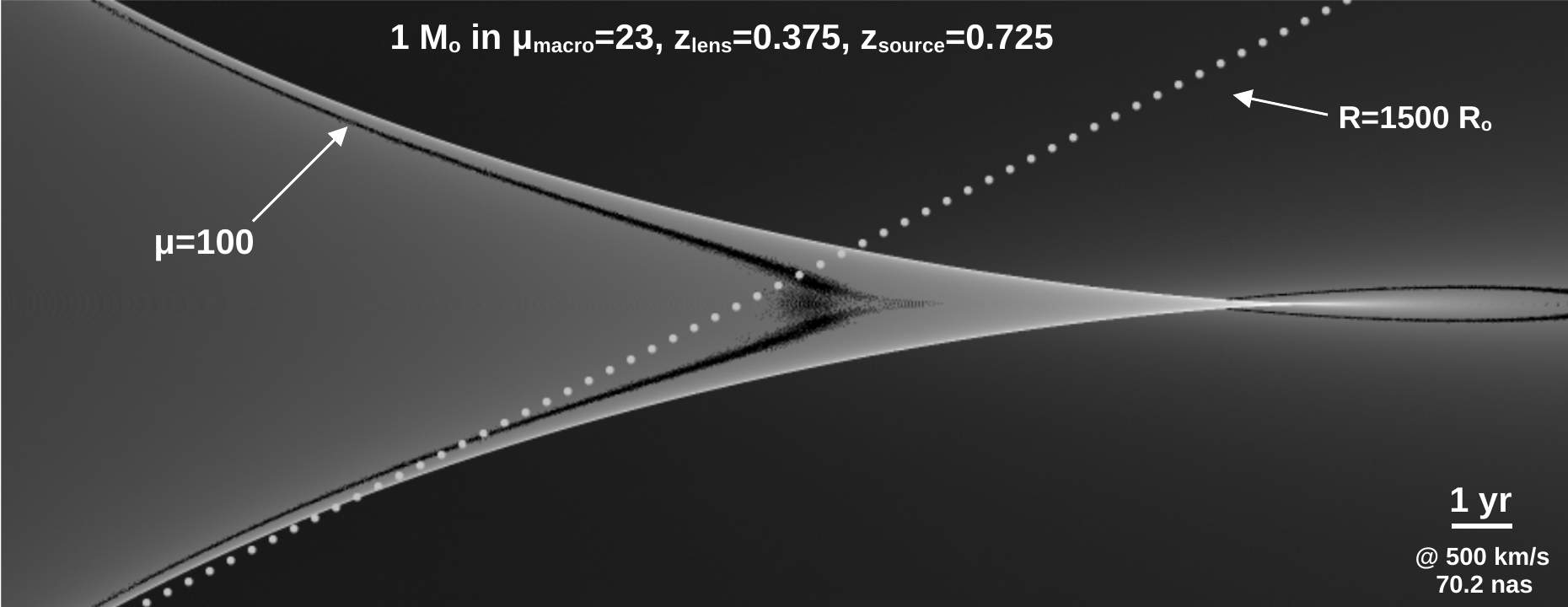}
      \caption{Cusp region around a microcaustic with $1\, \Msun$ in a macromodel with magnification $\mu=23$. The gray scale shows the logarithm of the magnification. The region near the caustic with magnification $\sim 100$ is marked in black. The white bar in the bottom right is the distance moved by a background star at $z=0.725$ in 1\,yr when the relative velocity is 500\,km\,s$^{-1}$. The maximum magnification is 3700 at the tip of the cusp. The pixel size is $320\,\Rsun$, so a supergiant star with diameter $80\,\Rsun$ would reach twice this magnification at the maximum, or $\mu_{\rm max}^{\rm tip}\approx 7400$. At the fold caustics, the maximum magnification for a  $80\, \Rsun$ diameter star would be smaller, $\mu_{\rm max}^{\rm fold}\approx 1000$. The straight line with dots shows the track of a hypothetical star moving across the caustic. The size of the dot corresponds to a star with radius $1500\, \Rsun$.}
         \label{Fig_1MsunMicrocaustic}
\end{figure}

At $\mu=100$, a star with absolute magnitude $-7$ would still be undetected within the far and near regions (apparent magnitude 31.24 for $\mu=100$). A few of these undetected stars will be close enough to a microcaustic in the source plane. Approximately half of these stars will be moving away from the microcaustic and hence remain undetected in the near future, but the other half will be moving closer to the microcaustic and become brighter over time. At magnification $\mu=1000$, the same star with absolute magnitude $-7$ can now be detected easily with {\it JWST} in exposures of 1\,hr. For a background star near a microcaustic, the time it takes to move from $\mu=100$ to $\mu=1000$ depends on several factors such as the mass of the microlens, the macromodel magnification, the relative velocity between the background star and the microcaustic, the direction of motion relative to the microcaustic, and the point of crossing of the microcaustic. To get a sense of this timescale, we assume a microlens with mass $1\,\Msun$, the same macromodel magnification of the fiducial model $\mu_{\rm 1m}=23$, and a relative velocity of $v\, {\rm cos}\,\alpha=500$\,km\,s$^{-1}$, where $\alpha$ is the angle between the direction of motion and the microcaustic. In this situation, when the background star is at magnification $\mu=100$, the microcaustic is $\sim 0.1\,\mu$as away as shown in Figure~\ref{Fig_1MsunMicrocaustic}.  

At a velocity of $v\, {\rm cos}\,\alpha =500$\,km\,s$^{-1}$, a star with absolute magnitude $-7$, within a microcusp and with magnification $\mu\approx 100$, would take $\sim 1$\,yr to reach the caustic and become detectable. At the tip of the cusp, the magnification for a supergiant star with diameter $80\, \Rsun$ reaches a maximum of $\sim 7500$, and $\sim 1000$ at the fold caustics. At this velocity, this maximum magnification can be maintained for $\sim 1.5$\,days \cite{Miralda1991}, after which the magnification will drop to a factor of a few and the star will no longer be detectable. Approximately 1/3 of the background stars in the far region with absolute magnitude $-7$, and near a microcaustic with magnification $\mu \approx 100$, will move toward the caustic and become detectable after 0.5--1\,yr of observation (or similarly, they are detectable now and will disappear behind the microcaustic after 1\,yr or move away from the caustic).
The exact same reasoning applies for the more numerous stars with double the magnification, or absolute magnitude $-7 +2.5\log_{10}(2)=-6.25$, but in an area 8 times smaller in the source plane ($A/d\mu \propto \mu^{-3}$) and with magnification $\mu>200$. But in this case, the distance to the microcaustic is four times smaller, so the cadence should be higher in order to detect these stars before they cross the caustic and become undetectable again. 

\begin{figure} 
   \includegraphics[width=9.0cm]{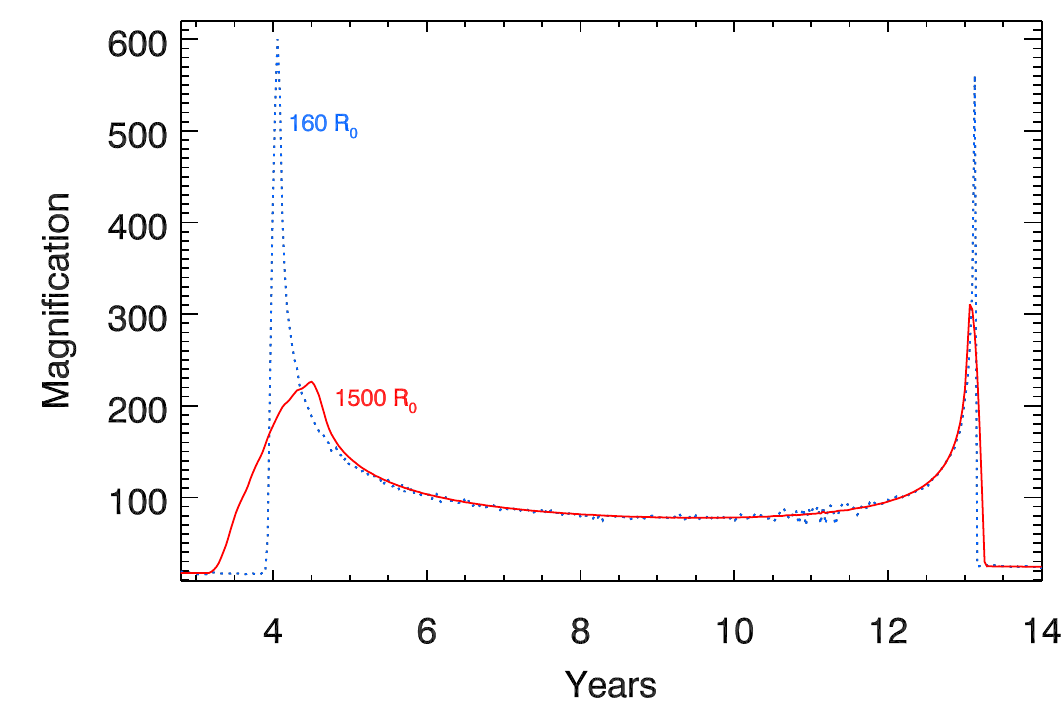}
      \caption{Light curve of a star moving at 500\,km\,s$^{-1}$ along the track shown in Figure~\ref{Fig_1MsunMicrocaustic}. The blue dotted line corresponds to a star with the same diameter as the pixel of the simulation ($320\,\Rsun$), while the red solid line corresponds to a much larger and more luminous star with  $R=1500\, \Rsun$. The star crosses the entire caustic region in $\sim 9$\,yr, with the first peak grazing the caustic and producing a wide peak.
         }
         \label{Fig_LightCurve}
\end{figure}

One factor to keep in mind is the very large radius of the most luminous RSGs, that can reach radii of $\sim 1500\, \Rsun$ \citep{Meynet2015}. Since the maximum magnification is lower for larger stars, these very luminous RSGs would generally have smaller magnification factors, typically a few hundred as shown in Figure~\ref{Fig_LightCurve}. When addressing the detectability of these stars, care needs to be taken to account for the smaller maximum magnification of large RSGs. 
Interestingly, the large radii of massive RSGs should correlate with where they are observed. If an RSG has a radius of $1500\,\Rsun$ and it can only be magnified by factors of a few hundred in the far region, only those RSGs exceeding $10^5\, \Lsun$ can be observed at $z=0.725$ during peak magnification. Less luminous RSGs of similar size can still be observed in the near region since multiple microcaustics overlap and the magnification can be $\mu>2000$ in this case (or 2.5\,mag deeper). The most luminous RSG is expected to have a maximum luminosity close to the HD limit, $L_{\rm max} \approx 5\times 10^5\, \Lsun$ or absolute magnitude $\sim -9.5$, which at magnification 100 can still be detected at $z=0.725$ with apparent magnitude $\sim 28.7$, so anywhere in the far region provided they are close enough to a microlens or millilens. 
BSGs, on the other hand, can be even more luminous and smaller ($R\lesssim 25\, \Rsun$), so during a microlensing event they can be magnified by factors $\mu>1000$ in the far region, allowing us to see the fainter (but more magnified) BSGs or the brighter but with more moderate and more likely magnification factors ($\mu<50$).

\subsection{Critical magnification}\label{subsec_mucrit}
The definition of the far-to-near ratio depends on the ratio of areas near to and far from the cluster CC, which in turn depends on our choice for the critical magnification. Taking a larger $\mu_{\rm crit}$ by a factor of 3 would lower the orange (and green) bands in Figure~\ref{Fig_Area_vs_NGC} and Figure~\ref{Fig_Area_vs_micro} by a factor of 9, but it would also lower the solid lines by the same factor of 9, leaving our conclusions unchanged. Hence, our conclusions are independent of the particular value of $\mu_{\rm crit}$.  
There is, however, a relatively small dependence on $\mu_{\rm crit}$ impacting our findings. The results in  Figure~\ref{Fig_Area_vs_NGC}  and Figure~\ref{Fig_Area_vs_micro} are normalized to the area in the far region, which for  $\mu_{\rm crit}=100$ represents 83.4\% of the area covered by the Dragon arc. Lowering $\mu_{\rm crit}$ would reduce this fraction of area where far events from millilenses can take place. The dependence of the fraction of area in the near region scales almost linearly with $\mu_{\rm crit}$, so we can approximate this as $F_{\rm far}\approx 1 - 16.7/\mu_{\rm crit}$, where 16.7 is the fraction of area in the near region when $\mu_{macro}=100$.  For example, from this law we find $F=66.6\%$ and $F=91.6\%$ for $\mu_{\rm crit}=50$ and $\mu_{\rm crit}=200$, respectively, while from the lens model we find  $69.3\%$ and $93.2\%$. Hence, in the conservative case where $\mu_{\rm crit}=50$ (this would require a very high $\Sigma_{\rm crit}>100\, \Msunpc^{-2}$), we find that the lines in Figures~\ref{Fig_Area_vs_NGC} and \ref{Fig_Area_vs_micro} would be corrected by a factor of $66.6/83.4=0.8$, while for $\mu_{\rm crit}=200$, the same curves would move upward by 
a small amount $91.6/83.4=1.1$,
leaving our results virtually unchanged.

\subsection{Slope of the lensing potential} 
Related to the previous point, another source of uncertainty impacting our results is the specific properties of the cluster lens model, in particular the slope of the lensing potential. The macromodel magnification enters in Equation~\ref{Eq_AGTmu} quadratically. If the macromodel magnification in the far region is 3 times larger, this would increase the amplitude of the solid lines in Figure~\ref{Fig_Area_vs_NGC} by a factor of 9, bringing the prediction from millilenses and the observation to better agreement. 
The median and mean  magnifications of the WSLAP+ model in the far region of the Dragon arc are 19.2 and 27.7, respectively. Increasing the magnification by a factor of 3 would bring the most common values of $\mu_{\rm 1m}$ in Equation~\ref{Eq_AGTmu} close to the value of $\mu_{\rm crit}$, resulting in a very uniform distribution of microlensing events along the Dragon arc. 
Such a lens model would require a very shallow lensing potential, possibly in conflict with lensing constraints. Comparing our lens model magnification in the far region with the predicted magnification from the lens models in the same region of \cite{Keith2024}, we find that on average those models predict $24\% \pm22\%$ more magnification than our lens model in the far region. 
Based on this, and taking the upper limit ($46\%$ increase), we expect the solid colored lines in Figure~\ref{Fig_Area_vs_NGC} to increase by a factor of $\sim 2.1$, still insufficient to explain the low rate of predicted events.

\subsection{Number density of millilenses}  
\label{sect_Satellites}
$N$-body simulations show a tight correlation between the virial mass of the cluster and the number of GCs, $M_{\rm vir}=5\times 10^9\, \Msun \times N_{\rm GC}$ \citep{Burkert2020,Valenzuela2021}. The galaxy clusters in which transient stars have been found are all very massive, with virial masses $\sim 10^{15}\, \Msun$. Hence, we expect  $\sim 2\times 10^5$ GCs in each of these clusters.
It is difficult to estimate with precision the expected number density of GCs (detected and nondetected) at the positions of the transients, but we can get an order-of-magnitude estimate and see if it is in agreement with the observed densities in nearby clusters. 

If we assume that the distribution of GCs follows a cored isothermal profile, then the number density of GCs falls with distance to the center as $\sim (R_c+R)^{-1}$. Assuming all $\sim 2\times 10^5$ GCs in the cluster are within a radius of 1\,Mpc with this profile, we find that the number densities at $R=50$\,kpc and $R=70$\,kpc vary between $\sim 0.76$ and $\sim 1.13$ per kpc$^2$ when the core radius ($R_c$) varies between 0 and 10\,kpc. This is within a factor of 2 of what was assumed in Figure~\ref{Fig_Area_vs_NGC}, and hence this higher estimate of the number density is still insufficient to explain with millilenses the anomalously high observed ratio of far-to-near microlensing events. Modifying the radial profile to a steeper one with number density scaling as $(R_c+R)^{-2}$ [similar to what would be expected if the distribution of GCs follows the profile from  \cite{NFW1996}, and for the most favorable scenario with $R_c=0$, the number density increases to $\sim 1.6$ and $\sim 2.9$ GCs per kpc$^2$ at 70\,kpc and 50\,kpc (respectively) from the center of the halo. This is still insufficient, since it would place the ratio of far-to-near events one order of magnitude below the observed rate.

The mass function in Figure~\ref{Fig_GC_massfunction} excludes halos more massive than a few times $10^7\, \Msun$. Naturally, we expect halos in this mass range to still contribute as millilenses, but it is unclear how many of those exist, since their potential must be shallow enough to not contain dense concentrations of stars in their central regions, hence evading direct detection. 
 
Nevertheless, dwarf galaxies, or small satellites in general, are expected to be numerous in cluster environments and introduce perturbations in the magnification in the far region (and also in the near region). From the lensing point of view, their cored structures and relatively low mass make many of them subcritical (that is, they do not produce CCs). However, the fraction of critical to subcritical halos remains unknown in cluster environments, so it is difficult to accurately predict their contribution to the probability of high magnification. Even if they do not reach criticality, at distances of $\sim 1''$ from the cluster CC, the mass associated with a satellite may be enough to alter the inverse of the magnification $\mu_{\rm 2m}^{-1}=(1-\kappa_{\rm 2m})^2-\gamma_{\rm 2m}^2$ near the satellite, and bring it close to the small value needed for microlensing events to be maximized, $|\mu_{\rm 2m}^{-1}|\approx 10^{-2}$. 


\begin{figure} 
   \includegraphics[width=9.0cm]{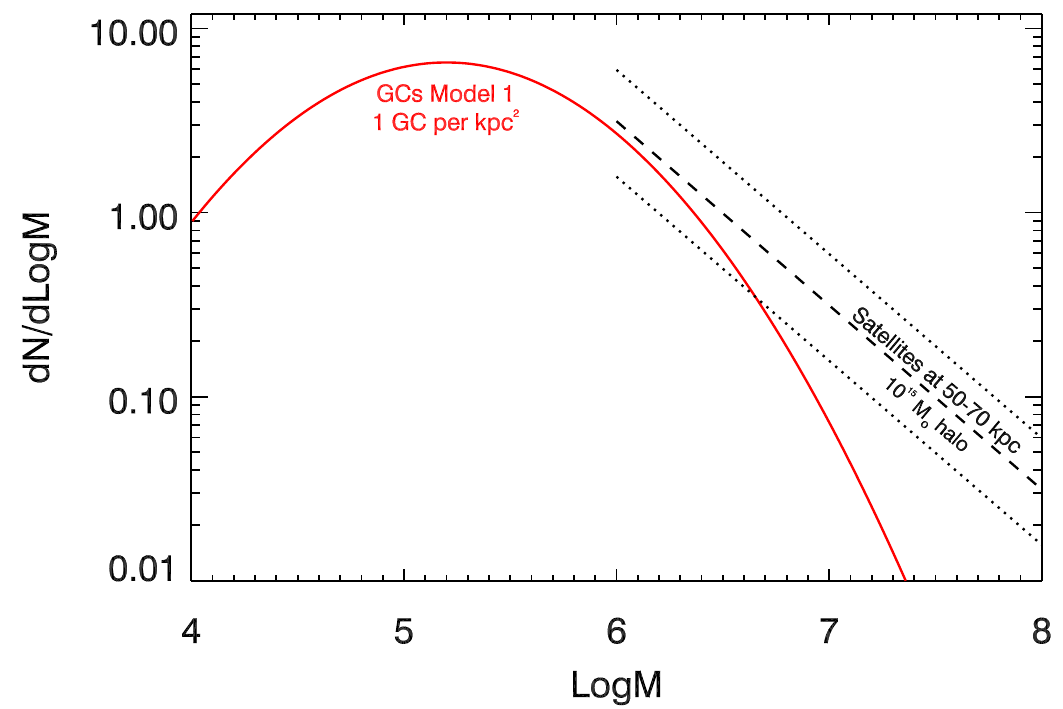}
      \caption{Comparison of mass functions of GCs and satellites. The red solid line shows the total number of GCs in the area corresponding to the Dragon arc from the GC mass function, after normalizing it to a density of 1 GCs per kpc$^2$. The black dashed line shows the mass function from $N$-body simulations of a population of satellites in a cluster with $M_{\rm vir}=10^{15}\, \Msun$ and computed in a similar area at distances between 50 and 70\,kpc from the center of the halo (same distance of the Dragon arc). The dotted lines shows the dispersion in the number of satellites from 16 different realizations.  In most realizations, no halos more massive than $\sim 10^8\, \Msun$ are found within the region considered. The total mass from the GC mass function is $4.01\times10^8\, \Msun$ and the total mass from the satellite mass function is $3.07\times10^8\, \Msun$. Finally, the sum of mass from the GC mass function up to $10^6\, \Msun$ and the mass from the satellite mass function above $10^6\, \Msun$ is $5.01\times10^8\, \Msun$.
         }
         \label{Fig_Satellites}
\end{figure}

Here we rely on results from pseudo-analytical realizations based on state-of-the-art recipes calibrated using numerical $N$-body simulations to assess the contribution from undetected satellite galaxies. 
We employed a sample of 16 very-high-resolution realizations using the \texttt{MOKA} algorithm \citep{giocoli12a,giocoli16b} 
assuming a mass of $10^{15}\,M_{\odot}$ at $z=0.37$. The cluster-size halo is populated by Monte Carlo sampling 
the subhalo mass function measured by \citep{giocoli10a} and extrapolating it to $1.5 \times 10^5\,M_{\odot}$. The halos have 
a triaxial model \citep{jing02,despali14} and subhalos are spatially distributed as calibrated by \citet{gao04}.

The resulting mass function from satellites is shown in Figure~\ref{Fig_Satellites}, where we compare it with the mass function of GCs used in our main result (Model 1). For this comparison, we have computed the total number of GCs in the area of the Dragon arc ($\sim 960$\,kpc$^2$) and assuming the number density of GCs is 1 per kpc$^2$, roughly the upper limit of the vertical blue band in Figure~\ref{Fig_Area_vs_NGC}. The mass function of satellites is also normalized to the same area covered by the Dragon arc, and it corresponds to the abundance of satellites with masses larger than $10^6\,M_{\odot}$ found at distances between 50 and 70\,kpc from the cluster center in simulated clusters at  $z\approx 0.37$, and with virial mass $10^{15}\, M_{\odot}$.  Rather than repeating the calculation we did for GCs, we can estimate the contribution from these satellites in the most favorable situation. We assume that all satellites are supercritical and compact enough so they contribute to the magnification similarly to GCs --- that is, they follow the scaling of Equation~\ref{Eq_AGTmu}. This is an optimistic scenario because a fraction of these satellites will be subcritical. In fact, it is unlikely that a large fraction of them are supercritical since this would imply they have dense detectable cores, and none is clearly observed as a resolved source in the vicinity of the Dragon arc. Nevertheless, under the ideal assumption above, the upper-limit contribution to the area in Equation~\ref{Eq_Atot} from the satellites should be proportional to their total integrated mass. We compute this mass from the dashed-line model shown in Figure~\ref{Fig_Satellites} and find a total mass of $3.07\times10^8\, \Msun$ in the area occupied by the Dragon arc. Repeating the same calculation for the red solid curve in Figure~\ref{Fig_Satellites}, we find that the GCs contribute $4.01\times10^8\, \Msun$ in the same area, a factor of 4.7  more. Considering instead a combined mass function composed of the red curve up to $10^6\, \Msun$ and the dashed black line above this mass, the total mass is $5.01\times10^8\, \Msun$, or $25\%$ more than the GC contribution. Translating these numbers into Figure~\ref{Fig_Area_vs_NGC}, the red curve (Model 1) would move upward by only a  factor of 1.25. Hence, even in the most optimistic case in which satellites are very compact and supercritical, the contribution from the undetected satellite galaxies is relatively minor.

\subsection{Substructure along the line of sight}  
\label{sect_LOS}
So far we have assumed the milli-lens substructure capable of promoting a micro-lensing event to detectability lies inside the virial radius of the cluster. However, CDM also predictions a sizeable population of halos along the line of sight (LOS), which can also contribute to the lensing perturbations \citep[e.g.][]{Gilman19}. On average, we expect substructures to contribute to the surface mass density an amount similar to the contribution from the mean density of the universe, $\bar{\rho}=\Omega_m\times\rho_{crit}$. In CDM, the overwhelming majority of dark matter halos on the relevant mass scales are subcritical, and the lensing effects of these objects drop when placed close to the observer and source.  Therefore, we consider LOS contributions from the redshift range $0.15 < z < 0.5$, assuming the source is the Dragon galaxy (at $z=0.725$). In this interval, the critical density of the universe is, on average, 1.41 times higher than at $z=0$, $\rho_{crit}(z=0)=2.77\times10^{11}\times h^2\, \Msun {\rm Mpc}^{-3}$. Projecting along the line of sight (1268 Mpc comoving), we then get an average contribution of $\Sigma_{LOS} \sim 242.7\, \Msunpc^{-2}$. Unlike subhalos of the cluster, dark matter halos outside the cluster environment are not affected by tidal forces. 

To assess the contribution from halos outside the virial radius of the cluster, we calculate the expected number of dark matter halos in the mass range $10^5$--$10^9$\,M$_{\odot}$ using the mass function model presented by \citet{Sheth1999}. In addition to the halos drawn from the Sheth-Tormen mass function, we account for correlated structure around the cluster. The mass of the cluster $10^{14}$--$10^{15}$ \,M$_{\odot}$ causes a local enhancement to the density field that increases the number of dark matter halos within $\sim 5\,\rm{Mpc^{-1}}$; these objects are effectively at the same redshift as the cluster itself, but are not inside the virial radius or even necessarily bound to the cluster potential, and thus they are typically not included in satellite mass functions. We model the local enhancement through the two-halo term \citep{Gilman19}, with the additional correction proposed by \citep{Lazar2021}. The halos along the line of sight, including those corresponding to correlated structure around the cluster, contribute 100--200\,M$_{\odot}\, \rm{pc^{-2}}$, depending on the assumed virial mass of the cluster. This constitutes a significant contribution that can potentially impact the results discussed in previous paragraphs. From Figure~\ref{Fig_BetaSigma_2D}, an increase of this magnitude in $\Sigma$ would make the amount of substructure in the far region very large, $\Sigma_{\rm Tot}= \Sigma_{*}+\Sigma_{\rm LOS}\approx 150\, \Msunpc^{-2}$, which would imply a ratio of events of $\sim 1$ in the far and near regions, and in conflict with the observed ratio of $\sim 2.5$. Alternatively, for a fixed ratio of events $\sim 2.5$, a larger value for $\Sigma_{\rm Tot}$ would imply a large value of $\beta\gtrsim 3$. 
However, in CDM dark matter halos on these scales are predicted to be subcritical for lensing, even when placed on top of a cluster convergence map, and thus they contribute subdominantly to the lensing magnification perturbations required to boost the signal from microlenses. We will be able to revisit this topic with future measurements of more events in the Dragon arc that will enable measurements of $\beta$ directly, and provide a better estimation of the ratio of events in the near and far regions, resulting in a constraint on the contribution from microlenses and millilenses in the lens plane and in the LOS to $\Sigma_{\rm Tot}$.

\begin{figure*} 
   \includegraphics[width=18.0cm]{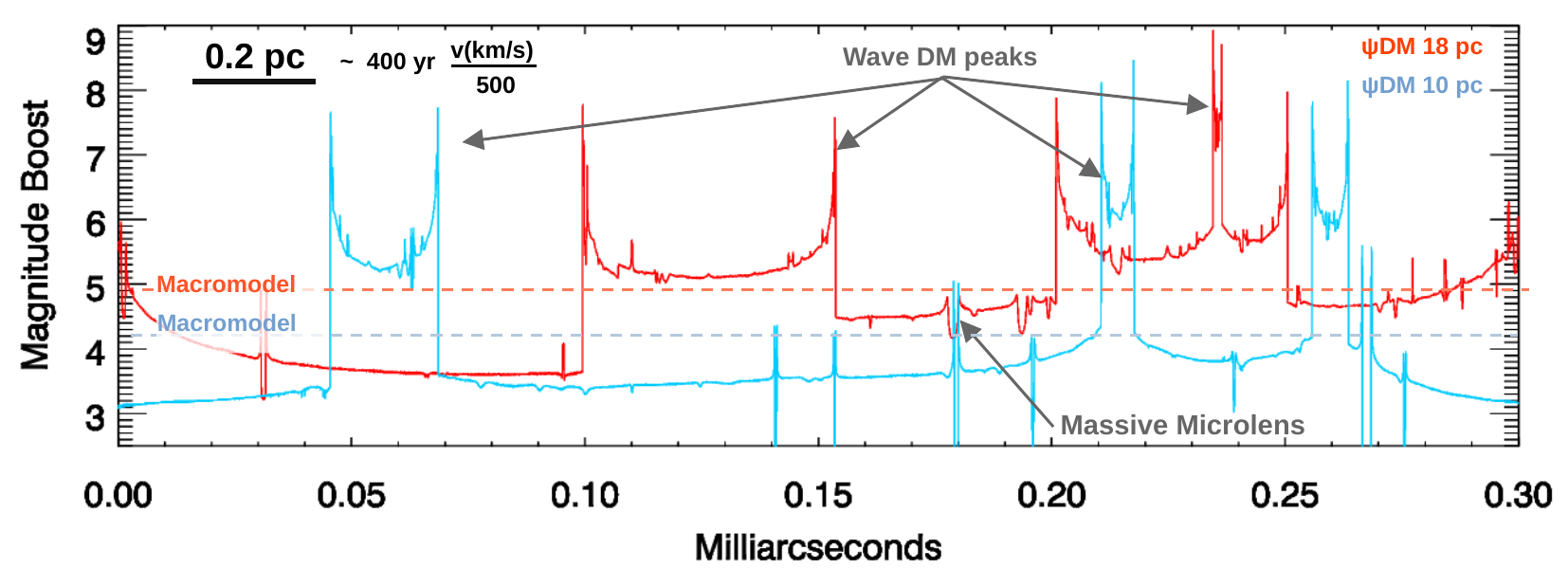}
      \caption{Simulated magnification (in magnitudes) for a source traversing a caustic region. The lens model includes $\psi$DM with de Broglie wavelength 18\,pc (red) or 10\,pc (blue). Two macromodel magnifications are shown: $\mu_{\rm macro} \approx 100$ (red) and $\mu_{\rm macro} \approx 50$ (blue). The effects of different lens components are marked. For this figure, the number density of microlenses was reduced to just $1\, \Msunpc^{-2}$ in order to better show the effect of $\psi$DM. Even at moderate values of $\mu_{\rm macro}$ (blue), the effects of $\psi$DM are very different from standard CDM expectations.
        }
         \label{Fig_wDM}
\end{figure*}

\subsection{Presence of hyperluminous stars}
The events found in the Dragon galaxy by the Flashlights program were observed in only one filter, so unfortunately we lack color information to assess whether these events could be LBVs at moderate magnifications or microlensing events of the much more abundant but fainter supergiant stars.

In low-redshift arcs such as the Dragon arc, hyperluminous stars with absolute magnitude $-11$ or brighter could be seen anywhere around the arc in regions with moderate magnification factors of $\mu_{\rm 1m}> 20$. Such high luminosity can be reached, for instance, during an outburst of an LBV reaching an absolute magnitude of $-11$ \citep{Weis2020}, or apparent magnitude $\sim 29$ with magnification $\mu \approx 20$. Outbursts as luminous as $-14$\,mag have been recorded \citep{Pastorello2010}, or even brighter for the so-called ``supernova impostors" that can be as luminous as supernovae \citep[e.g.,][]{Kilpatrick2018}. These superluminous LBVs are exceedingly rare and can be detected anywhere at this redshift without the help of magnification, so we do not consider them here. 
In the case of  typical LBV outbursts, they would be more likely detected in the far region since this corresponds to a larger area in the source plane.

In the Dragon arc, \cite{Keith2024} estimate $\sim 3$ LBVs should be present. 
We can independently estimate the number of LBVs if we assume the Dragon galaxy contains a similar number of LBVs as the number found in our neighborhood. The number of LBVs with absolute magnitude brighter than $-10$ found in the Milky Way plus LMC plus SMC is $N_{\rm LBV}\approx 100$ \citep{Humphreys1979,Hamann2006,Crowther2010,Hainich2014}. 
To see one of these stars without help from microlensing or millilensing, the magnification needs to be $\mu_{\rm 1m} \approx 50$.  
Only a portion of the Dragon arc has magnification $> 50$. We can estimate this by multiplying the upper bound of the orange region in Figure~\ref{Fig_Area_vs_NGC} by a factor of 4 --- that is, $\sim  5$ kpc$^2$ in the Dragon arc are magnified by a factor of 50 or more. 
This corresponds to a fraction of $6.3\% \times (R_{\rm gal}/5 {\rm kpc})^2$, 
where we have adopted the estimated radius of the Dragon galaxy from our lens model, $R_{\rm gal}\approx 5$\,kpc. The number of expected LBVs in this area is then $N\approx 6$, close to the estimate from \cite{Keith2024}. Most of these LBVs likely will be in a quiescent phase and hence not detected as transients when comparing observations separated by 1\,yr (or $\sim 0.5$\,yr in the source frame), but for observations at two epochs separated by several years, a significant fraction of them will show measurable changes in flux and be identified as transients. We conclude that some of the events found in the far region of the Dragon arc may be LBVs, but without color information we cannot confirm this hypothesis.

At higher redshifts, larger magnification factors are needed to see outbursting LBVs, so the expectation in this case is to see mostly genuine microlensing events in the far region. An example (but also an exception) is Godzilla, a star which is believed to be an outbursting LBV at $z=2.37$ with at least 5 counterimages \citep{DiegoGodzilla}, all of them (but one) undetected at macromodel magnification $\mu_{\rm 1m}\approx 100$ (the example), but interestingly with one being detected thanks to the magnification boost provided by a millilens ($\mu_{\rm 2m}\gtrsim 2000$) (the exception). \cite{DiegoGodzilla} estimate that at any given point $\sim 30$ extremely magnified LBVs (EMBLVs) at  $1<z<3$ and with magnification $> 1000$ should be detectable in the sky and with apparent magnitudes as bright as 24. Large-scale high-cadence surveys such as LSST can reveal them and complete a census of EMBLVs up to $z\approx 3$.

\subsection{Alternatives to $\Lambda$CDM}\label{subsec_wDM}
Although the combination of standard microlensing (that is, not involving a millilenses) and LBVs offers the simplest explanation for the high ratio of events found in the far region, it is interesting to consider other scenarios in which dark matter physics, or various baryonic effects \citep[e.g.][]{Ragagnin2024}, alter the properties of halos. For example, warm DM models predict less substructure on subgalactic scales. Surviving halos in warm DM have lower concentrations than their CDM counterparts, and therefore have a suppressed lensing efficiency, lowering the contribution from millilenses to the lensing probability. On the other hand, self-interacting DM can cause halos to undergo core collapse, a process that dramatically raising their central density, potentially to a degree that causes them to become super-critical for lensing \citep{Gilman21}. Alternatively, wave dark matter, ($\psi$DM) is expected to increase the magnification in the far region. In this model, DM has density fluctuations at scales given by the de Broglie wavelength and the halo mass \citep{Schive2016}, from the dependence on momentum:
\begin{equation}
\lambda_{dB}=15\,\left( \frac{10^{-22}\, {\rm eV}}{m_{\psi}} \right) \left( \frac{10^{15}\, \Msun}{M_{\rm cluster}}\right)^{1/3}\, \, {\rm pc}\, ,
\end{equation}
where $m_{\psi}$ is the mass of the ultralight axion-like particle (ALP). 
For masses $m_{\psi} \approx 10^{-22}$\,eV and a $10^{15}\, \Msun$ cluster, this scale corresponds to 3\,mas in the lens plane. This pervasive interference substructure causes the CC to become corrugated on the de Broglie scale \citep{Chan2020,Laroche+22,Amruth2023}, and increasingly so for more massive halos, with many detached islands where the magnification diverges at relatively large offsets from the cluster CC \citep{Amruth2023,Laroche+22,Powell2023}. 

Similarly to microlenses, $\psi$DM fluctuations are ubiquitous across the lens plane, and as in the case of microlenses and millilenses, these fluctuations get amplified near the CC by the macromodel. 
In Figure~\ref{Fig_wDM} we show the effect of $\psi$DM over a small region in the observer plane. For this particular case the source is at $z=1$, but the effect would be very similar for $z=0.725$. The simulation of $\psi$DM follows \cite{Amruth2023}, and for this particular case the value of $\Sigma_{*}$ has been decreased to $\Sigma_{*}=1\, \Msunpc^{-2}$, to better appreciate the $\psi$DM effect. Two models for $\psi$DM are considered with $\lambda_{\rm dB}=10$\,pc and $\lambda_{\rm dB}=18$\,pc, for cluster-scale lenses. 
We also consider two macromodel magnifications. As shown in the figure, $\psi$DM introduces perturbations in the magnification pattern in the source plane at the sub-milliarcsecond level. This scale is similar to the scale of the caustics from the GCs considered earlier, and is consistent with results from analyses of multiply-imaged quasars in which the effect from $\psi$DM is shown to be comparable to the effect of population of halo millilenses \citep{Laroche+22}. Interestingly, smaller masses for the ALP (blue curve) result in more pronounced effects but over a smaller region. In both cases, a significant portion of the source plane can attain sufficiently high magnifications so the critical magnification is reached, maximizing the probability for microlensing effects to take place.



The de Broglie wavelength (hence the mass of the ALP) and the macromodel magnification determine the type of object that can exhibit different magnifications. For instance, in the $\lambda_{\rm dB}=10$\,pc model and $\mu_{\rm 1m}=50$ in Figure~\ref{Fig_wDM}, the scale of the object needs to be typically larger than $\sim 0.1$\,pc in order to be insensitive to $\psi$DM fluctuations, while for the $\lambda_{\rm dB}=18$\,pc and $\mu_{\rm 1m}=50$ model the source needs to be larger than 0.5\,pc in order to not exhibit asymmetric fluxes. Future observations of the Dragon arc will reveal additional microlensing events, which are expected to form clusters of microlensing events around the strongest fluctuations in the boson field.

\section{Conclusions}\label{sect_concl}
\label{Sect_Conclusions}
We study the 3M-lensing effect from the combination of a macromodel, a millilens, and microlenses. The possibility is considered that microlensing events found at relatively large distances from the cluster CC in the Dragon arc, or far region, are  aligned with millilenses in the lens plane that increase the probability of microlensing. We study the scaling of the area above certain magnification (or lensing probability) near a  millilens, with the mass of the millilens and macromodel magnification, with and without adding microlenses. Near the cusps of millilenses, this probability scales with the mass of the millilens, and microlenses play a minor role. We consider a realistic population of millilenses and model their mass function with a log-normal function, then compute the total area in the far region of the source plane associated with this population of millilenses that has magnification greater than some critical value. We find that the contribution to this area from millilenses is less than the contribution from the far more numerous microlenses elsewhere in the source plane.  Hence, the addition of millilenses does not appreciably increase the expected rate of microlensing events far from the critical curve (which is given mostly by the more numerous microlenses). Other factors, such as the presence of LBVs, also contribute  to the number of transient events in the far region, especially in lensed galaxies at low redshift where LBVs  can be detected even at modest magnification factors. 

We pay special attention to the spatial distribution of microlensing events and find that the number density of microlensing events also depends on the exponent of the LF, $\rho({\mu},\beta) \propto \mu^{\beta-2}$. We make the analogy of traditional photographic-plate imaging and identify two regimes: (i) positive-imaging regime when $\beta>2$ and the number density of microlensing events is higher around massive substructures (high $\mu$), and (ii) negative-imaging regime when $\beta<2$ where microlensing and microlensing events have smaller number densities at the position of massive substructures (also high $\mu$). 

We discuss the intimate relationship between the abundance of DTM stars and the number of observed microlensing events where the second is proportional to the former. We demonstrate, both analytically and with Monte Carlo simulations, how the number density of DTM stars shows a strong dependence on the LF and the macromodel magnification. Once the population of DTM stars has been established (from the LF and the macromodel magnification), the problem of estimating the number of microlensing events can be reduced to studying a population of DTM stars as they move across the web of microcaustics, where the later depends not only on the amount of substructure (microlenses and millilenses), but also on the macromodel magnification. 
We use the observed density of events in the far and near regions of the Dragon arc to derive the slope of the LF, finding that a steep LF with $\beta=2.55^{+0.72}_{-0.56}$ is consistent with the observations. 
Variation of the LF along the lensed Dragon Arc or absorption by dust are not considered in this work but they should add an additional element of uncertainty in the results.
With future data, one can measure the slope $\beta$ directly from the observed LF and confront it with our estimate of $\beta=2.55^{+0.72}_{-0.56}$ derived from the spatial distribution of the number density of microlensing events.  

We derive a relation between the slope of the LF, $\beta$, the amount of substructure, $\Sigma$, and the ratio of observed microlensing events in the near and far regions, $N_{\rm near}/N_{\rm far}$. 
We estimate the amount of substructure along the line of sight and, from the relation between $\beta$, $\Sigma$, and $N_{\rm near}/N_{\rm far}$, we argue that most of this substructure should be in the form of subcritical halos. Otherwise, the inferred values of $\beta$ would be very high.

Small substructures in the far region of the CC can be mapped (imaged) by measuring this number density of microlensing events, which should correlate with the location of millilensing substructures. The clustering may also reveal a non-uniform distribution of the background stellar population that can equally show clustering. Repeated observations of the same arc may be and a detailed analysis of the photometry (or spectra if available) may be needed in order to clearly distinguish between the two scenarios. We apply this technique to two microlensing events forming a pair of local high density, and under the assumption of a uniform distribution of the background stars, find that if this peak in the density of microlensing events is due to a substructure, its mass is $\sim 1.3\times10^8\, \Msun$ within its Einstein radius. This technique shall open a new window to map the distribution of mass on scales of milliarcseconds, including perturbations in the DM field. As an illustration, we consider the case of $\psi$DM and argue that this type of model can be proven with repeated observations of low-redshift caustic-crossing arcs, such as the Dragon arc, thereby greatly increasing the statistics on the spatial distribution of microlensing events and revealing the hidden nature of DM at subarcsecond scales.  \\

At the time of submission of this paper, new JWST observations of this arc have revealed more than 40 microlensing candidates in the near and far regions of the Dragon arc. Most of these events are suspected to be due to  RSG stars at $z=0.725$. These events are presented in \cite{Fudamoto2024}. A detailed analysis of these new events will be the subject of a future paper. 

\begin{acknowledgements}
 We thank Ian Smail for useful comments and suggestion. This research was  supported by NASA/HST grants GO-15936, GO-16278, and GO-16729 from STScI, which is operated by the Association of Universities for Research in Astronomy, Inc. under NASA contract NAS5-26555.  
 J.M.D. acknowledges the support of project PID2022-138896NB-C51 (MCIU/AEI/MINECO/FEDER, UE) Ministerio de Ciencia, Investigaci\'on y Universidades.  
 C.G. is grateful for the support from INAF theory Grant 2022: Illuminating Dark Matter using Weak Lensing by Cluster Satellites, PI Carlo Giocoli.
 A.V.F. was supported by the Christopher R. Redlich Fund and many individual donors.
\end{acknowledgements}

\bibliographystyle{aa} 
\bibliography{MyBiblio} 

\begin{appendix}

\section{Lens Model}
Details of the algorithm are provided by \cite{Diego2005,Diego2007,Diego2016}. This modelling technique has been applied successfully to several clusters observed with {\it HST} and {\it JWST} \citep{Diego2005,Diego2007,Diego2016,Diego2023Gordo,Diego2023MACS0416}. 

The model for A370 is derived using 32 lensed galaxies with spectroscopic redshifts, and producing over 90 multiple images, or constraints. The model is derived as part of the Beyond the Ultradeep Frontier Fields and Legacy
Observations (BUFFALO) project \citep[GO-15117, PIs Steinhardt \& Jauzac;][]{Steinhardt2020}, and also incorporates  information from weak lensing measured with {\it HST} images. Details of the dataset are given by \cite{Niemiec2023}. 

This model incorporates all member galaxies detected by {\it HST} near the Dragon arc, so it includes all relevant deflectors at galactic scales and above.  
The CC predicted by our lens model in the Dragon arc is shown in Figure~\ref{Fig_Dragon}. 
For this work we are interested in the area in the source plane with magnification $\mu>\mu_{\rm crit}$ from millilenses that are in regions of the lens plane where $\mu_{\rm 1m}<\mu_{\rm crit}$ (or far region). We are also interested in a similar area in the source plane but from regions in the lens plane near the cluster CC where the macromodel alone can provide $\mu_{\rm 1m}>\mu_{\rm crit}$ needed for the probability of microlensing to be high enough. More precisely, we are interested in the ratio of the two areas, since this ratio will essentially correlate with the ratio of events found near the CC and far from the CC. The use of a different lens model should have a relatively small impact on our conclusions provided these lens models include all member galaxies near the critical region, since these member galaxies can alter the position of the CCs. The ratio of events should then remain more or less constant for most models, with a relatively small dependence on the slope of the lensing potential. This difference in slopes can account for a factor of $\sim 2$ in the ratio of areas and hence on the ratio of events between different lens models. A level of uncertainty of a factor of $\sim 2$ should be kept in mind owing to uncertainty in the macro galaxy cluster model. 

\begin{figure} 
   \includegraphics[width=9.0cm]{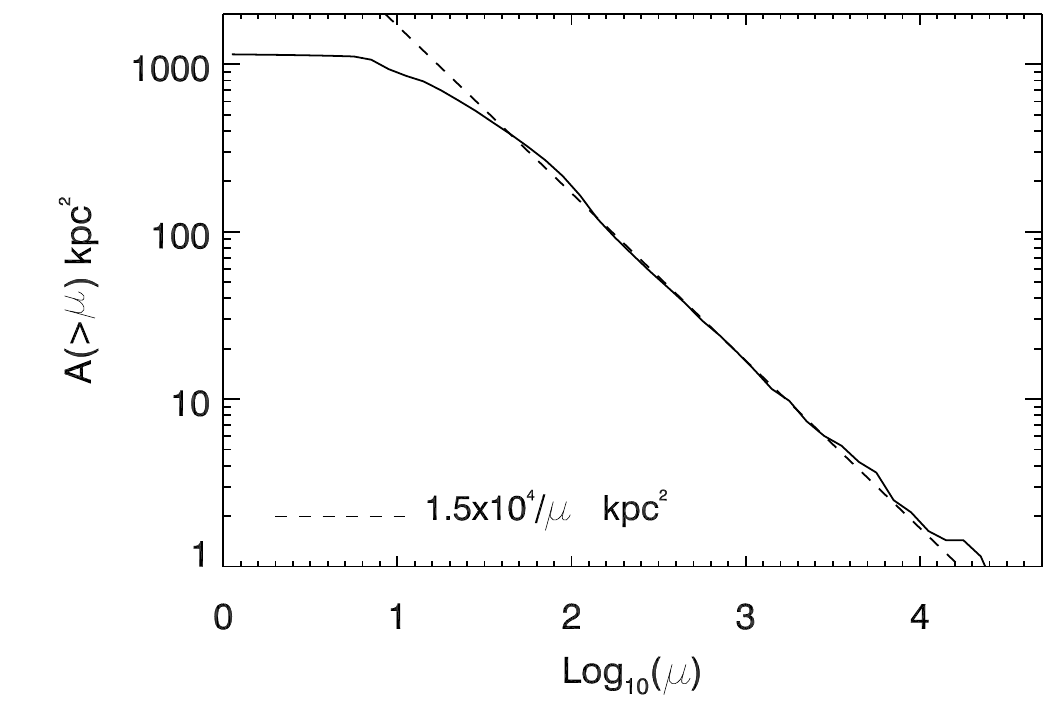}
      \caption{Area above a certain magnification in the Dragon arc.
               The solid line shows the area with magnification $>\mu$ computed in the image plane and in the region occupied by the Dragon arc. The dashed line is the simple power-law fit $A(>\mu) = 1.5\times 10^4/\mu$ in kpc$^2$.
         }
         \label{Fig_AGTmuDragon}
\end{figure}

The area in the image plane above a magnification $\mu$ computed in the region of the Dragon arc  is shown in Figure~\ref{Fig_AGTmuDragon}. As expected, this area scales as the canonical $1/\mu$ scaling law. 
Above $\mu=100$ there are 190\,kpc$^2$ in the image plane. Dividing by $\mu=100$, this corresponds to 1.9\,kpc$^2$ in the source plane, and correcting for the multiplicity factor 2 (for every counterimage with  magnification $\mu=100$ in the image plane, there is another on the other side of the CC with similar magnification), we arrive at 0.95\,kpc$^2$, setting the upper boundary of the  orange region in Figure~\ref{Fig_Area_vs_NGC}.

\end{appendix}

\end{document}